\documentclass[aps,prm,reprint,twocolumn,superscriptaddress,showpacs,amssymb]{revtex4-1} 
\usepackage[english]{babel}
\usepackage{graphics}
\usepackage{graphicx}
\usepackage{epsfig}
\usepackage{amssymb}
\usepackage{dcolumn}
\usepackage{bm}
\usepackage{color}
\usepackage{lineno}
\usepackage{sidecap}
\usepackage{verbatim}  
\usepackage{vector}  
\usepackage{wrapfig}
\usepackage{enumerate}
\usepackage{sidecap}
\usepackage{amsmath}
\usepackage{hyperref}

\begin{document}

\title{Temperature independent band structure of WTe$_2$ as observed from ARPES}

\author{S. Thirupathaiah}
\email{t.setti@sscu.iisc.ernet.in}
\affiliation{Solid State and Structural Chemistry Unit, Indian Institute of Science, Bangalore, Karnataka, 560012, India.}
\author{Rajveer Jha}
\affiliation{CCNH, Universidade Federal do ABC (UFABC), Santo Andre, SP, 09210-580 Brazil.}
\author{Banabir Pal}
\affiliation{Solid State and Structural Chemistry Unit, Indian Institute of Science, Bangalore, Karnataka, 560012, India.}
\author{J. S. Matias}
\affiliation{CCNH, Universidade Federal do ABC (UFABC), Santo Andre, SP, 09210-580 Brazil.}
\author{P. Kumar Das}
\affiliation{CNR-IOM, TASC Laboratory AREA Science Park-Basovizza, 34149 Trieste, Italy.}
\affiliation{International Centre for Theoretical Physics, Strada Costiera 11, 34100 Trieste, Italy.}
\author{I. Vobornik}
\affiliation{CNR-IOM, TASC Laboratory AREA Science Park-Basovizza, 34149 Trieste, Italy.}
\author{R. A. Ribeiro }
\affiliation{CCNH, Universidade Federal do ABC (UFABC), Santo Andre, SP, 09210-580 Brazil.}
\author{D. D. Sarma}
\affiliation{Solid State and Structural Chemistry Unit, Indian Institute of Science, Bangalore, Karnataka, 560012, India.}

\date{\today}

\begin{abstract}
Extremely large magnetoresistance (XMR),  observed in  transition metal dichalcogendies, WTe$_2$,   has attracted recently a great deal of research interests as it shows no sign of saturation up to the magnetic field as high as 60 T, in addition to the presence of type-II Weyl fermions. Currently, there has been a lot of discussion on the role of band structure changes on the temperature dependent XMR in this compound. In this contribution, we study the band structure of WTe$_2$  using angle-resolved photoemission spectroscopy (ARPES) and first-principle calculations to demonstrate that the temperature dependent band structure has no substantial effect on the temperature dependent XMR as our measurements do not show band structure changes on increasing the sample temperature between 20 and 130 K. We further observe an electronlike surface state, dispersing in such a way that it connects the top of bulk holelike band to the bottom of bulk electronlike band. Interestingly, similar to bulk states, the surface state is also mostly intact with the sample temperature. Our results provide invaluable information in shaping the mechanism of temperature dependent XMR in WTe$_2$.
\end{abstract}

\maketitle

\subsection{Introduction}

Materials showing extremely large magnetoresistance (XMR) have potential applications in spintronics. Among them, the semimetals, WTe$_2$ and MoTe$_2$, have attracted  a great deal of research interests recently as they show nonsaturating extremely large MR~\cite{Ali2014, Ali2015,Zhou2016} even at 60 T of applied field in addition to the prediction of Weyl-nodes~\cite{Soluyanov2015, Sun2015a}. So far there exists several theories for the XMR observed in metals. While the metal-insulator transition in presence of magnetic field is mostly applied mechanism of XMR in metals~\cite{Khveshchenko2001, Kopelevich2006, Wang2014, Zhao2015, Xiang2015}, for some specific compounds such as Ag$_{2+\delta}$Te/Se~\cite{Xu1997, Lee2002}, graphene~\cite{Singh2012},  topological insulators Bi$_2$Te$_3$ and Bi$_2$Se$_3$ ~\cite{Qu2010, Wang2012b}, Dirac semimetals Cd$_3$As$_2$~\cite{Liang2014, Feng2015} and Na$_3$Bi~\cite{Kushwaha2015} and type-I Weyl semimetals TaAs~\cite{Huang2015,Zhang2015a},  NbAs~\cite{Ghimire2015} and NbP~\cite{Shekhar2015, Wang2016a} that are showing linear field dependent XMR these are the massless Dirac fermions near the Fermi level which cause this effect~\cite{Abrikosov1998, Abrikosov2003}. On the other hand,  the compounds such as type-II Weyl semimetals WTe$_2$ and MoTe$_2$~\cite{Ali2014, Zhou2016}, LaSb~\cite{Zeng2016} and ZrSiS~\cite{Lv2016} that are showing quadratic field dependent  XMR,  it was predicted that the charge compensation causes the effect.  Nevertheless, the theory of charge compensation for the nonsatuarting XMR in type-II Weyl semimetals is also debated~\cite{Zandt2007, Thirupathaiah2017}.

An intriguing property of XMR materials is the turn-on temperature, below which the resistivity increases rapidly for a non-zero magnetic field~\cite{Ali2014}. And this turn-on temperature increases with field, suggesting band structure changes with magnetic field but not with temperature alone. On the other hand, earlier ARPES studies reported temperature dependent band structure in WTe$_2$ ~\cite{Pletikosic2014, Wu2015} measured in the absence of magnetic field and intuitively correlated the temperature dependent band structure with the XMR. Technically however, it is highly unlikely that WTe$_2$ shows a temperature dependent band structure as it neither shows structural nor electronic phase transition down to the lowest possible temperature from the room temperature~\cite{Kabashima1966,Ali2014}. In agreement with this view, a recent band structure study using first-principle calculations with the inclusions of temperature effect suggests for no dramatic changes in the band structure of WTe$_2$ below 300 K~\cite{Liu2017}.  Hence, it is interesting to study the band structure of WTe$_2$ as a function of temperature to reveal the discrepancies between the experiment and the theory and to further understand the origin of XMR.

\begin{figure*}[htbp]
	\centering
		\includegraphics[width=0.98\textwidth]{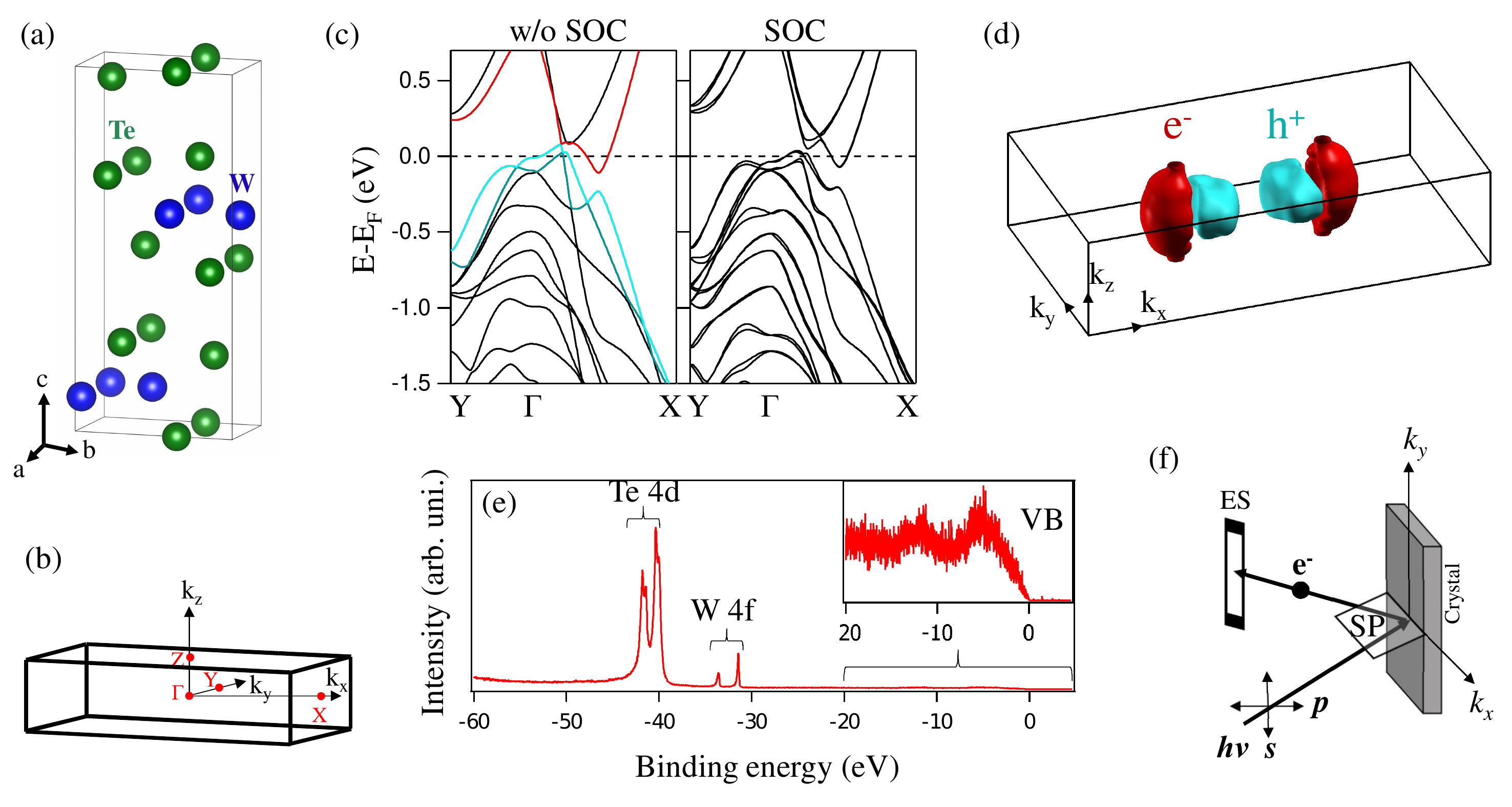}
	\caption{(a) Orthorhombic crystal structure of WTe$_2$. (b) 3D view of the bulk Brillouin zone on which the high symmetry points are located. (c) Band structure  of WTe$_2$ from the DFT calculations performed without and with spin-orbit coupling (SOC). (d) 3D view of the Fermi surface map derived without SOC.  (e) Angle integrated photoemission spectra with the core-level energy positions labeled and the zoomed-in valence spectra is shown in the inset. (f) Schematic of a typical measuring geometry in which the $s$- and $p$-plane polarized lights are deﬁned with respect to the analyzer entrance slit (ES) and the scattering plane (SP).}
	\label{1}
\end{figure*}

  In this contribution, we report the electronic structure studies of WTe$_2$ using high-resolution angle-resolved photoemission spectroscopy  and first-principle calculations. We noticed two holelike and two electronlike pockets in the Fermi surface map. In addition to this we detected an electronlike surface state, dispersing in such a way that it connects the top of bulk hole pocket to the bottom of bulk electron pocket. These results are in good agrement with our band structure calculations and with the previous photoemission studies on this system~\cite{Pletikosic2014, Wu2015, Jiang2015,Bruno2016, Wu2016, Wang2016b,Das2016, Belopolski2016,Feng2016, Sanchez-Barriga2016a}. Orbital-resolved band structure calculations suggest that near the Fermi level the bands are mainly composed by W $5d$ and Te $5p$. We further noticed strong hybridization between these two orbital characters. Our experimental results further suggest that the band structure of this compound is temperature independent within the range of measured temperature (20-130 K). Interestingly, the electronlike surface state also persistent throughout the sample temperature treatment.  These results are in stark contrast to some of the pervious ARPES reports~\cite{Pletikosic2014, Wu2015} where it was suggested that the band structure of WTe$_2$ is highly sensitive to the temperature. Here, we discuss the plausible reasons for the discrepancies between our results and Refs.~\onlinecite{Pletikosic2014, Wu2015} and the implications of our experimental findings in understanding the temperature dependent XMR of this compound.

\subsection{Experimental and Band structure calculation details}

High quality single crystals of stoichiometric WTe$_2$ were grown using the self-flux method at Universidade Federal do ABC (UFABC), Brazil as discussed in Ref.~\onlinecite{Zhou2016}.  The crystals have a platelet-like shape with shiny surface. ARPES measurements were performed at the APE beamline in Elettra Synchrotron,  Trieste equipped with a Scienta DA30 deflection analyzer. The angular resolution was set at $0.3^\circ$ and the overall energy resolution was set at 25 meV.  Samples were cleaved $\textit{in situ}$ at a temperature of 20 K and the chamber vacuum was better than $5\times10^{-11}$ mbar. During the measurements the sample temperature was varied between 20 and 130 K.

Band structure calculations are performed on the orthorhombic crystal structure of WTe$_2$~\cite{Brown1966},  having the lattice constants $a$ = 3.496 \AA,~ $b$ = 6.282 \AA,~ and $c$ = 14.07 \AA~ using density functional theory (DFT) within the generalized gradient approximation (GGA) of Perdew, Burke and Ernzerhof (PBE) exchange and correlation potential~\cite{Perdew1996} as implemented in the Quantum Espresso simulation package~\cite{QE-2009}. Norm conserving scalar relativistic and fully relativistic pseudopotentials are used to perform the calculations without spin-orbit coupling (SOC) and with SOC, respectively. The electronic wavefunction is expanded using plane waves up to a cutoff energy of 50 Ry (680 eV). Brillouin zone sampling is done over a (24$\times$14$\times$6) Monkhorst-Pack $k$-grid. During the calculation we have fixed the experimentally obtained lattice parameters but relaxed the internal atomic coordinates.

\begin{figure*}
	\centering
		\includegraphics[width=0.98\textwidth]{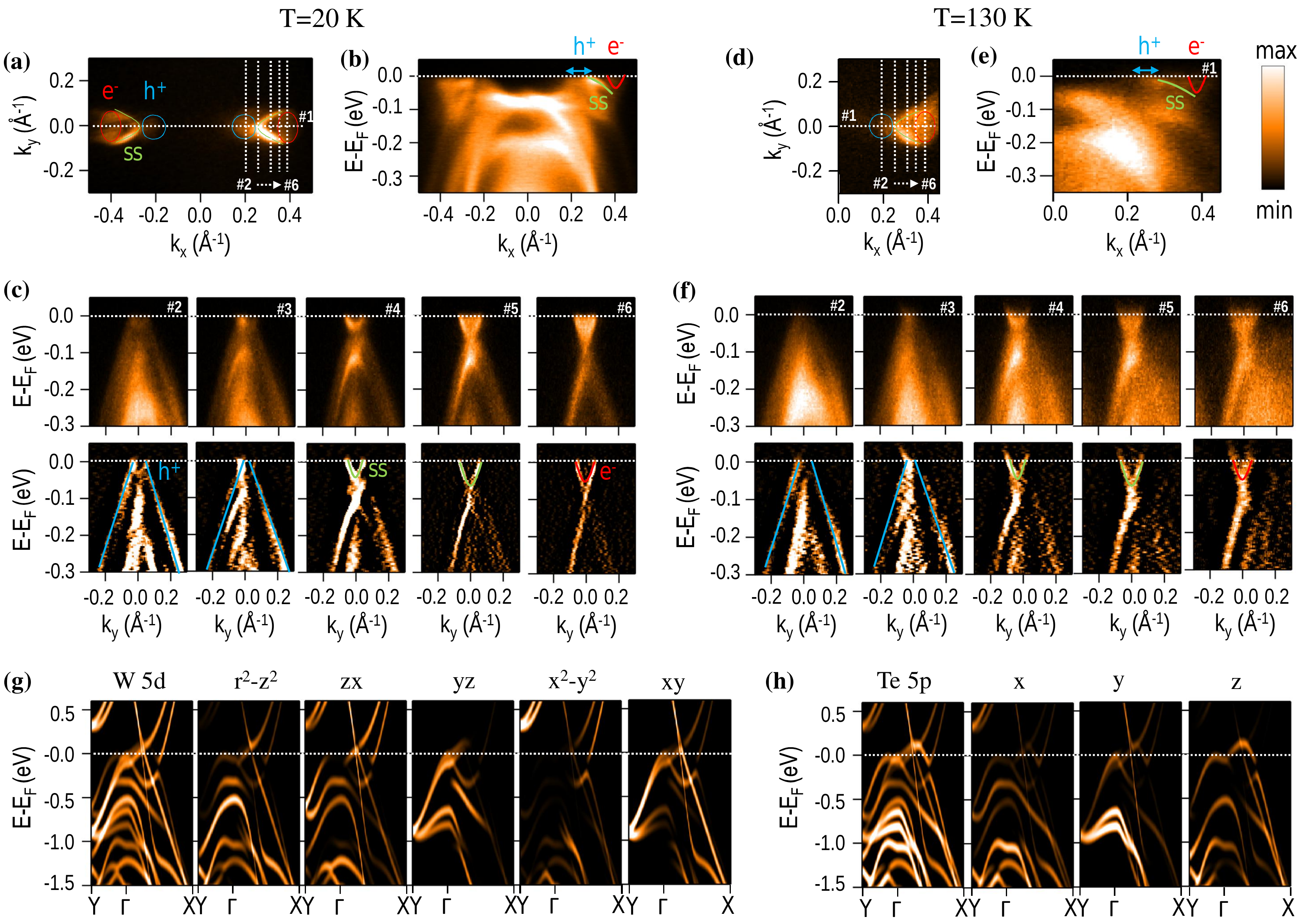}
	\caption{ ARPES data of WTe$_2$. The data are measured using $p$-polarized light with a photon energy of h$\nu$=20 eV. The data shown in (a)-(c) are measured at a sample temperature of 20 K. (a) depicts Fermi surface (FS) map. Hole and electron pockets are schematically shown by blue and red color solid contours (contribution from bulk) and green color contour is the Fermi arc from the surface. (b) shows energy distribution map (EDM) taken along the cut \#1 as shown on FS map. Top panels in (c) show EDMs taken along the cuts \#2 $-$ \#6 from left to right, respectively. Bottom panels in (c) are the respective second derivatives of the EDMs shown in the top panels. On the EDMs in (b) and (c) the band dispersions are schematically shown.  (d)-(f) depict similar data of (a)-(c) except that these are measured at 130 K. (g) and (h) depict the orbital-resolved band structure from the calculations plotted for W 5$d$ and Te 5$p$ orbital characters, respectively.}
	\label{2}
\end{figure*}



\subsection{Results and Discussions}

Fig.~\ref{1}(c) depicts energy ($E$)-momentum ($k$) plot obtained from first-principle calculations without and with spin-orbit coupling. In the band structure obtained without SOC,  we showed holelike bands in the blue color and the electronlike bands in the red color that are contributing to the Fermi surface topology. 3D view of the Fermi surface map derived without SOC is shown in Fig.~\ref{1}(d) where we could find two hole and two electron pockets shown by red and blue colors, respectively. Note here that with SOC, one should find 4 hole and 4 electron pockets contributing to the Fermi surface map. As can be seen further from Fig.~\ref{1}(d), the electron pockets show a strong $k_z$ warping in going from $\Gamma-Z$, while the hole pockets show negligible $k_z$ warping and also the pocket terminates abruptly at halfway between $\Gamma$ and $Z$. Therefore, no hole pockets are present at the $Z$-point.  Thus, the calculated band structure suggests that WTe$_2$ is a 3D electronic system though it has a layered crystal structure~\cite{Brown1966, Ali2014}. Also it was recently suggested using ARPES~\cite{DiSante2017} that WTe$_2$ is indeed a 3D electron system. Our band structure calculations are in very good agrement with the previous reports~\cite{Dawson1987, Augustin2000, Ali2014, Soluyanov2015, Das2016}.  Angle integrated photoemission spectra is shown in Figure~\ref{1}(e). The spectra is taken with a photon energy of 100 eV. In Fig.~\ref{1}(e) , the core-level energy positions of Te $4d$ and W $4f$ are identified and the zoomed-in valance band spectra is shown in the inset. Apart from the core-levels of W and Te, we did not find any impurity peaks in Fig.~\ref{1}(e)


In Figure~\ref{2}, we show ARPES data of WTe$_2$ measured with a photon energy of 20 eV using $p$-polarized light. The data shown in Figs.~\ref{2} (a)-(c) are measured at a sample temperature of 20 K. Using inner potential of 11.5 eV~\cite{Pletikosic2014}, we calculated that 20 eV photon energy extracts the bands from $k_z$ = 6 $\pi/c$ plane. From the Fermi surface map shown in Fig.~\ref{2}(a),  we can identify two holelike and two electronlike pockets on either side of the $\Gamma$-point along the $k_x$ direction. As can be further seen from Fig.~\ref{2}(a) that the Fermi surface topology of these compounds is highly anisotropic, which means, the  spectral intensity distribution along the $k_x$ direction is entirely different from that of along the $k_y$ direction. This observation is in-line with anisotropy of the crystal structure as shown in Fig.~\ref{1}(a).  In addition to the bulk hole and electron pockets, we notice an electronlike Fermi arc connecting both the bulk hole and electron pockets as shown by a green line in Fig.~\ref{2}(a). This Fermi arc is ascribed to the presence of Weyl nodes in WTe$_2$~\cite{Jiang2015,Bruno2016, Wu2016, Wang2016b, Das2016, Belopolski2016,Feng2016, Sanchez-Barriga2016a, Crepaldi2017}.


To further elucidate the nature of band dispersions near the Fermi level ($E_F$),  we made cuts along the $k_x$ and $k_y$ directions as shown in Figs.~\ref{2} (b) and (c), respectively. The bottom panels in Fig.~\ref{2}(c) are the respective second derivatives of the top panels.  From Fig.~\ref{2}(b), the EDM cut taken in the $k_x$ direction,  one can notice that the electronlike surface state disperses in such a way that it connects bottom of the bulk electronlike band to the top of bulk holelike band. As the band structure of these compounds is complex near the Fermi level, difficult to disentangle the individual bands. Hence, we showed momentum range of holelike band dispersions on the EDM plot [see Fig.~\ref{2}(b)]. From the EDM cuts taken in the $k_y$ direction,  we identified bulk holelike [see cuts~\#2 and~\#3 in Fig.~\ref{2}(c)], surface electronlike [cuts~\#4 and~\#4], and bulk electronlike band dispersions [cut~\#6]. These observations are consistent with the existing ARPES reports on WTe$_2$~\cite{Bruno2016, Wu2016, Wang2016b, Belopolski2016,Feng2016, Sanchez-Barriga2016a}.

\begin{figure}
	\centering
		\includegraphics[width=0.5\textwidth]{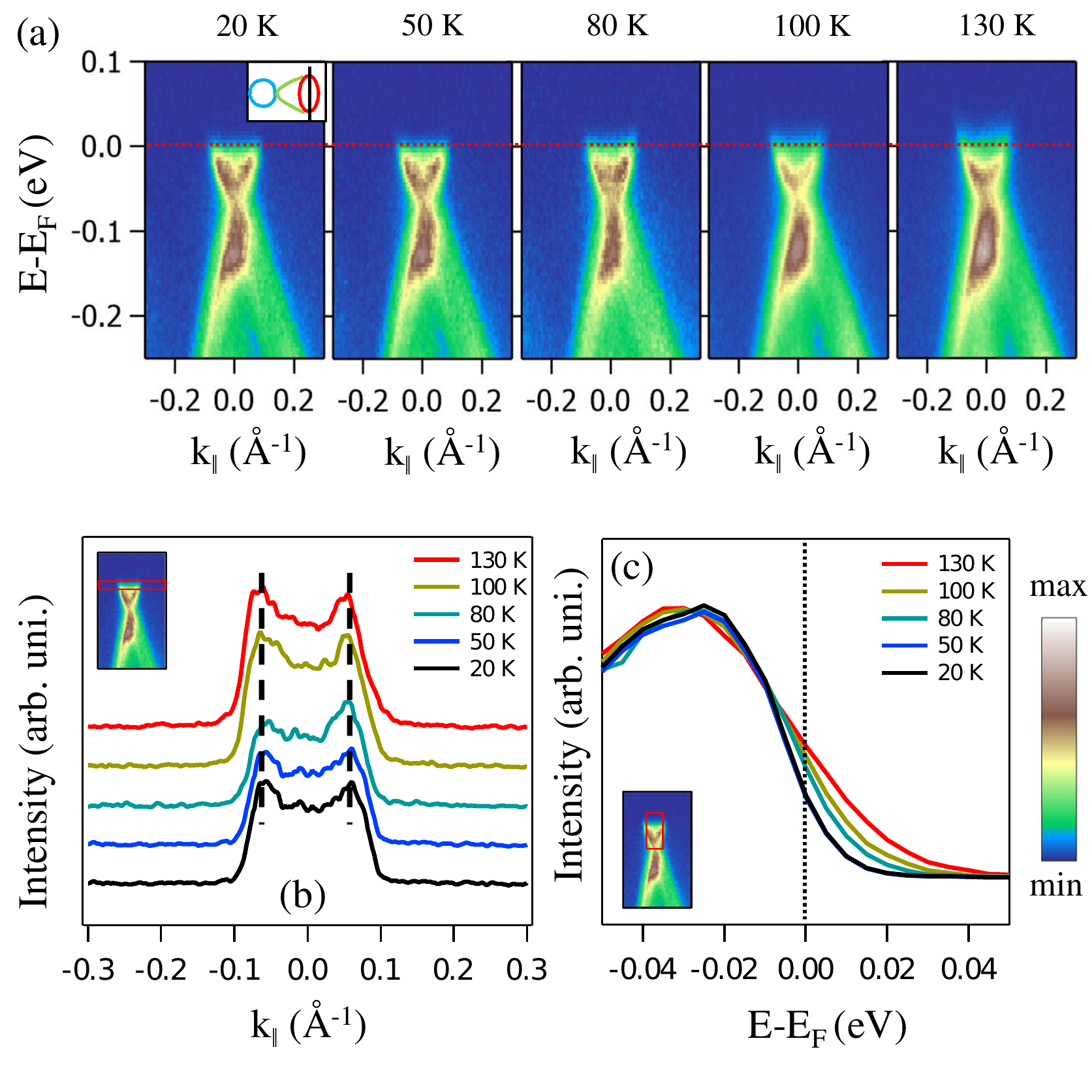}
	\caption{Temperature dependent ARPES data of WTe$_2$ measured with a photon energy of 20 eV using $p$-polarized light. Top panels in (a) depict EDMs showing the bulk electron pocket dispersion taken along the cut as shown in the inset. (b) Depicts momentum dispersive curves taken at the Fermi level with an energy integration window of 10 meV as shown in the inset. (c) Depicts energy dispersive curves taken from a momentum integration of of 0.1 $\AA^{-1}$ centered at $k_{||}$ = 0 as shown in the inset.}
	\label{3}
\end{figure}

 In Figures~\ref{2} (d)-(f) we show ARPES data of WTe$_2$ measured at a sample temperature of 130 K. All the data are measured using $p$ polarized light with a photon energy of 20 eV.  Fig.~\ref{2}(d) depicts Fermi surface map.  Fig.~\ref{3}(e) and (f) depict EDM cuts taken along the $k_x$ and $k_y$, respectively. The bottom panels in Fig.~\ref{2}(f) are the second derivatives of the top panels.  In Figs.~\ref{2}(g) and (h), we show the orbital-resolved band structure calculations for W $5d$ and Te $5p$. These calculations are obtained without including SOC interaction.  As can be seen from Figs.~\ref{2} (g) and (h), the band structure near the Fermi level is highly hybridized between W $5d$ and Te $5p$. Particularly, near the Fermi level its hard to disentangle the orbital characters of the hole and electron pockets. In some of the earlier reports,  4 hole and 4 electron pockets have been shown for WTe$_2$~\cite{Bruno2016, Wu2016, Wang2016b, Belopolski2016,Feng2016, Sanchez-Barriga2016a}, whereas in our ARPES study we could only identify two hole and two electron pockets. We think, the remaining hole and electron pockets can be detected using $s$-polarized light as these are composed by multiple orbital characters.

 On comparing Fig. ~\ref{2}(a), the Fermi surface map measured at 20 K,  with that of Fig.~\ref{2}(d), the Fermi surface map measured at 130 K, one can clearly notice that the the size of hole and electron pockets hardly changes with the temperature between 20 and 130 K. For a quantitative comparison of the band structure between these two temperatures, we further estimated the Fermi vectors of hole and electrons pockets. The estimated Fermi vectors for both bulk hole and electron pockets at 20 K are of approximately 0.041 $\AA^{-1}$ and 0.050 $\AA^{-1}$ , respectively. And at 130 K the Fermi vectors are of approximately 0.037 $\AA^{-1}$ and 0.055 $\AA^{-1}$ for the hole and electron pockets, respectively. These values are hinting for nearly equal number of electron and hole carriers at least from the the zone center, which is consistent with the theory of charge compensation for the XMR in WTe$_2$~\cite{Ali2014, Pletikosic2014}.

To further understand better the evolution of band structure in WTe$_2$ with temperature, we measured high quality data for the bulk electron pockets as a function of temperature as shown in Figure~\ref{3}.  Fig.~\ref{3}(a) depicts the temperature dependent EDMs, showing bulk electronlike band dispersion, taken along the cut shown by the black line on cartoon in the inset. Fig.~\ref{3}(b) depicts momentum dispersive curves (MDCs) from the EDMs shown in Fig.~\ref{3}(a), taken at the Fermi level by integrating over an energy window of 10 meV centered at the Fermi level (see inset). Fig.~\ref{3}(c) depicts energy dispersive curves (EDCs) Fig.~\ref{3}(a), extracted by integrating over a momentum window of 0.1 $\AA^{-1}$ centered at $k_{||}$ = 0 (see inset).

  As can be seen from  Figs.~\ref{3}(b) and ~\ref{3}(c), on increasing the sample temperature,  the size of electron pocket hardly changes between 20 and 130 K except for a typical thermal broadening. These results unambiguously demonstrate a temperature independent band structure in this compound. However, this observation is in stark  contrast to some of the earlier ARPES reports ~\cite{Pletikosic2014, Wu2015}, where it was suggested that with increase in temperature the size of electron and hole pockets increases and decreases, respectively. Moreover, a Lifshitz transition is noticed for the bulk hole pockets at a sample temperature of 160 K~\cite{Wu2015} as a result of monotonic decrease in the size hole pockets with the temperature.   However, from our studies we did not find any change in the size of hole and electron pockets between 20 and 130 K, thus, ruling out the possibility of a Lifshitz transition at 160 K.  Currently it is not clear to us why our temperature dependent measurements are in disagreement with Ref.~\onlinecite{Pletikosic2014, Wu2015}. Nevertheless, a quantitative comparison between our results and  the results from Ref.~\onlinecite{Pletikosic2014, Wu2015} reveals that our data is extracted from a different surface termination as we find the electronlike surface state that is not seen in their data. Therefore, the discrepancies between our results and Ref.~\onlinecite{Pletikosic2014, Wu2015} maybe related to the differing surface termination which may react differently with temperature as the Fermi surface of WTe$_2$ is significantly sensitive to the surface structure relaxation dynamics~\cite{Kawahara2017}. The present results are consistent with our recent temperature dependent ARPES studies on MoTe$_2$~\cite{Thirupathaiah2017}. Moreover, in MoTe$_2$ we did not find any cleavage dependent temperature effects on the band structure. Thus, it is important to check  whether a local band structure is playing any role here for the noticed discrepancies in the band structure of WTe$_2$.


  Though temperature dependent band structure of WTe$_2$ is a convincing theory to explain the temperature dependent XMR, it is hard to understand why the band structure is very sensitive to the temperature as this shows neither a structural nor an electronic phase transition below the room temperature~\cite{Kabashima1966,Ali2014}. Moreover, the quantum Hall measurements taken at 9 T suggest that the hole carrier density is intact below 150 K which is of $\approx$ 1 $\times$ 10$^{20}$ $cm^{-3}$, while it is the electron density ($\approx$ 5 $\times$ 10$^{20}$ $cm^{-3}$) that rapidly decreases below 50 K only to compensate with hole carriers below 5 K. In this regard, for any correlation between the band structure changes and the temperature dependent XMR, it is the size of electron pocket that has to be reduced with the temperature~\cite{Luo2015}. Though, constant hole density below 150 K as reported in Ref.~\onlinecite{Luo2015} is in-line with our results, rapid change in the electron density below 50 K is in contrast as we do not see changes neither in the size of hole pocket nor in the size of electron pocket between 20 and 130 K.  Therefore, our measurements suggest that the band structure changes may not be a reason for the temperature dependent XMR in WTe$_2$.  Our observations are further supporting a recent report~\cite{Wang2015b}, in which it was suggested that the temperature dependent band structure changes may not be having any role in the turn-on temperature of XMR materials.


\subsection{Conclusions}
In conclusion, we studied the low-energy band structure of WTe$_2$ by means of ARPES technique and first-principle calculations. We observe two holelike and two electronlike pockets from the Fermi surface map. In addition to the bulk bands,  we also detected electronlike surface state dispersing in such a way that it connects the top of bulk hole pocket to the bottom of bulk electron pocket.  Our orbital-resolved band structure calculations demonstrate a strong hybridization between W $5d$ and Te $5p$ near the Fermi level. In addition, we find from the calculations that particularly the holelike and electronlike pockets are formed from a combination of W $5d$ and Te $5f$ orbital characters. These results are in very good agrement with previous band structure studies using photoemission and first-principle calculations on this system.  Our experimental measurements further suggest that the band structure of WTe$_2$ is temperature independent between 20 and 130 K. Therefore, with the help of our experimental results we suggest that there is no direct relation between the band structure changes and the temperature dependent XMR in WTe$_2$. Thus, our present findings provide invaluable information in understanding the  mechanism of nonsaturating and temperature dependent XMR in WTe$_2$.

\subsection{Acknowledgements}

S.T.  acknowledges support by the Department of Science and Technology, India through the INSPIRE-Faculty program (Grant number: IFA14 PH-86). The authors acknowledge the financial support given for the measurements at Elettra Synchrotron under Indo-Italian (DST-ICTP) cooperation Program. R.A.R. acknowledges support by FAPESP (grant no. 2011/19924-2). This work has been partly performed in the framework of the nanoscience foundry and fine analysis (NFFA-MIUR Italy) project.

\bibliography{WTe2}

\begin{thebibliography}{51}%
\makeatletter
\providecommand \@ifxundefined [1]{%
 \@ifx{#1\undefined}
}%
\providecommand \@ifnum [1]{%
 \ifnum #1\expandafter \@firstoftwo
 \else \expandafter \@secondoftwo
 \fi
}%
\providecommand \@ifx [1]{%
 \ifx #1\expandafter \@firstoftwo
 \else \expandafter \@secondoftwo
 \fi
}%
\providecommand \natexlab [1]{#1}%
\providecommand \enquote  [1]{``#1''}%
\providecommand \bibnamefont  [1]{#1}%
\providecommand \bibfnamefont [1]{#1}%
\providecommand \citenamefont [1]{#1}%
\providecommand \href@noop [0]{\@secondoftwo}%
\providecommand \href [0]{\begingroup \@sanitize@url \@href}%
\providecommand \@href[1]{\@@startlink{#1}\@@href}%
\providecommand \@@href[1]{\endgroup#1\@@endlink}%
\providecommand \@sanitize@url [0]{\catcode `\\12\catcode `\$12\catcode
  `\&12\catcode `\#12\catcode `\^12\catcode `\_12\catcode `\%12\relax}%
\providecommand \@@startlink[1]{}%
\providecommand \@@endlink[0]{}%
\providecommand \url  [0]{\begingroup\@sanitize@url \@url }%
\providecommand \@url [1]{\endgroup\@href {#1}{\urlprefix }}%
\providecommand \urlprefix  [0]{URL }%
\providecommand \Eprint [0]{\href }%
\providecommand \doibase [0]{http://dx.doi.org/}%
\providecommand \selectlanguage [0]{\@gobble}%
\providecommand \bibinfo  [0]{\@secondoftwo}%
\providecommand \bibfield  [0]{\@secondoftwo}%
\providecommand \translation [1]{[#1]}%
\providecommand \BibitemOpen [0]{}%
\providecommand \bibitemStop [0]{}%
\providecommand \bibitemNoStop [0]{.\EOS\space}%
\providecommand \EOS [0]{\spacefactor3000\relax}%
\providecommand \BibitemShut  [1]{\csname bibitem#1\endcsname}%
\let\auto@bib@innerbib\@empty
\bibitem [{\citenamefont {Ali}\ \emph {et~al.}(2014)\citenamefont {Ali},
  \citenamefont {Xiong}, \citenamefont {Flynn}, \citenamefont {Tao},
  \citenamefont {Gibson}, \citenamefont {Schoop}, \citenamefont {Liang},
  \citenamefont {Haldolaarachchige}, \citenamefont {Hirschberger},
  \citenamefont {Ong},\ and\ \citenamefont {Cava}}]{Ali2014}%
  \BibitemOpen
  \bibfield  {author} {\bibinfo {author} {\bibfnamefont {M.~N.}\ \bibnamefont
  {Ali}}, \bibinfo {author} {\bibfnamefont {J.}~\bibnamefont {Xiong}}, \bibinfo
  {author} {\bibfnamefont {S.}~\bibnamefont {Flynn}}, \bibinfo {author}
  {\bibfnamefont {J.}~\bibnamefont {Tao}}, \bibinfo {author} {\bibfnamefont
  {Q.~D.}\ \bibnamefont {Gibson}}, \bibinfo {author} {\bibfnamefont {L.~M.}\
  \bibnamefont {Schoop}}, \bibinfo {author} {\bibfnamefont {T.}~\bibnamefont
  {Liang}}, \bibinfo {author} {\bibfnamefont {N.}~\bibnamefont
  {Haldolaarachchige}}, \bibinfo {author} {\bibfnamefont {M.}~\bibnamefont
  {Hirschberger}}, \bibinfo {author} {\bibfnamefont {N.~P.}\ \bibnamefont
  {Ong}}, \ and\ \bibinfo {author} {\bibfnamefont {R.~J.}\ \bibnamefont
  {Cava}},\ }\href {\doibase 10.1038/nature13763} {\bibfield  {journal}
  {\bibinfo  {journal} {Nature}\ }\textbf {\bibinfo {volume} {514}},\ \bibinfo
  {pages} {205} (\bibinfo {year} {2014})}\BibitemShut {NoStop}%
\bibitem [{\citenamefont {Ali}\ \emph {et~al.}(2015)\citenamefont {Ali},
  \citenamefont {Schoop}, \citenamefont {Xiong}, \citenamefont {Flynn},
  \citenamefont {Gibson}, \citenamefont {Hirschberger}, \citenamefont {Ong},\
  and\ \citenamefont {Cava}}]{Ali2015}%
  \BibitemOpen
  \bibfield  {author} {\bibinfo {author} {\bibfnamefont {M.~N.}\ \bibnamefont
  {Ali}}, \bibinfo {author} {\bibfnamefont {L.}~\bibnamefont {Schoop}},
  \bibinfo {author} {\bibfnamefont {J.}~\bibnamefont {Xiong}}, \bibinfo
  {author} {\bibfnamefont {S.}~\bibnamefont {Flynn}}, \bibinfo {author}
  {\bibfnamefont {Q.}~\bibnamefont {Gibson}}, \bibinfo {author} {\bibfnamefont
  {M.}~\bibnamefont {Hirschberger}}, \bibinfo {author} {\bibfnamefont {N.~P.}\
  \bibnamefont {Ong}}, \ and\ \bibinfo {author} {\bibfnamefont {R.~J.}\
  \bibnamefont {Cava}},\ }\href {\doibase 10.1209/0295-5075/110/67002}
  {\bibfield  {journal} {\bibinfo  {journal} {{EPL} (Europhysics Letters)}\
  }\textbf {\bibinfo {volume} {110}},\ \bibinfo {pages} {67002} (\bibinfo
  {year} {2015})}\BibitemShut {NoStop}%
\bibitem [{\citenamefont {Zhou}\ \emph {et~al.}(2016)\citenamefont {Zhou},
  \citenamefont {Rhodes}, \citenamefont {Zhang}, \citenamefont {Tang},
  \citenamefont {Sch\"onemann},\ and\ \citenamefont {Balicas}}]{Zhou2016}%
  \BibitemOpen
  \bibfield  {author} {\bibinfo {author} {\bibfnamefont {Q.}~\bibnamefont
  {Zhou}}, \bibinfo {author} {\bibfnamefont {D.}~\bibnamefont {Rhodes}},
  \bibinfo {author} {\bibfnamefont {Q.~R.}\ \bibnamefont {Zhang}}, \bibinfo
  {author} {\bibfnamefont {S.}~\bibnamefont {Tang}}, \bibinfo {author}
  {\bibfnamefont {R.}~\bibnamefont {Sch\"onemann}}, \ and\ \bibinfo {author}
  {\bibfnamefont {L.}~\bibnamefont {Balicas}},\ }\href {\doibase
  10.1103/PhysRevB.94.121101} {\bibfield  {journal} {\bibinfo  {journal} {Phys.
  Rev. B}\ }\textbf {\bibinfo {volume} {94}},\ \bibinfo {pages} {121101}
  (\bibinfo {year} {2016})}\BibitemShut {NoStop}%
\bibitem [{\citenamefont {Soluyanov}\ \emph {et~al.}(2015)\citenamefont
  {Soluyanov}, \citenamefont {Gresch}, \citenamefont {Wang}, \citenamefont
  {Wu}, \citenamefont {Troyer}, \citenamefont {Dai},\ and\ \citenamefont
  {Bernevig}}]{Soluyanov2015}%
  \BibitemOpen
  \bibfield  {author} {\bibinfo {author} {\bibfnamefont {A.~A.}\ \bibnamefont
  {Soluyanov}}, \bibinfo {author} {\bibfnamefont {D.}~\bibnamefont {Gresch}},
  \bibinfo {author} {\bibfnamefont {Z.}~\bibnamefont {Wang}}, \bibinfo {author}
  {\bibfnamefont {Q.}~\bibnamefont {Wu}}, \bibinfo {author} {\bibfnamefont
  {M.}~\bibnamefont {Troyer}}, \bibinfo {author} {\bibfnamefont
  {X.}~\bibnamefont {Dai}}, \ and\ \bibinfo {author} {\bibfnamefont {B.~A.}\
  \bibnamefont {Bernevig}},\ }\href {\doibase 10.1038/nature15768} {\bibfield
  {journal} {\bibinfo  {journal} {Nature}\ }\textbf {\bibinfo {volume} {527}},\
  \bibinfo {pages} {495} (\bibinfo {year} {2015})}\BibitemShut {NoStop}%
\bibitem [{\citenamefont {Sun}\ \emph {et~al.}(2015)\citenamefont {Sun},
  \citenamefont {Wu}, \citenamefont {Ali}, \citenamefont {Felser},\ and\
  \citenamefont {Yan}}]{Sun2015a}%
  \BibitemOpen
  \bibfield  {author} {\bibinfo {author} {\bibfnamefont {Y.}~\bibnamefont
  {Sun}}, \bibinfo {author} {\bibfnamefont {S.-C.}\ \bibnamefont {Wu}},
  \bibinfo {author} {\bibfnamefont {M.~N.}\ \bibnamefont {Ali}}, \bibinfo
  {author} {\bibfnamefont {C.}~\bibnamefont {Felser}}, \ and\ \bibinfo {author}
  {\bibfnamefont {B.}~\bibnamefont {Yan}},\ }\href {\doibase
  10.1103/PhysRevB.92.161107} {\bibfield  {journal} {\bibinfo  {journal} {Phys.
  Rev. B}\ }\textbf {\bibinfo {volume} {92}},\ \bibinfo {pages} {161107}
  (\bibinfo {year} {2015})}\BibitemShut {NoStop}%
\bibitem [{\citenamefont {Khveshchenko}(2001)}]{Khveshchenko2001}%
  \BibitemOpen
  \bibfield  {author} {\bibinfo {author} {\bibfnamefont {D.~V.}\ \bibnamefont
  {Khveshchenko}},\ }\href {\doibase 10.1103/PhysRevLett.87.206401} {\bibfield
  {journal} {\bibinfo  {journal} {Phys. Rev. Lett.}\ }\textbf {\bibinfo
  {volume} {87}},\ \bibinfo {pages} {206401} (\bibinfo {year}
  {2001})}\BibitemShut {NoStop}%
\bibitem [{\citenamefont {Kopelevich}\ \emph {et~al.}(2006)\citenamefont
  {Kopelevich}, \citenamefont {Pantoja}, \citenamefont {da~Silva},\ and\
  \citenamefont {Moehlecke}}]{Kopelevich2006}%
  \BibitemOpen
  \bibfield  {author} {\bibinfo {author} {\bibfnamefont {Y.}~\bibnamefont
  {Kopelevich}}, \bibinfo {author} {\bibfnamefont {J.~C.~M.}\ \bibnamefont
  {Pantoja}}, \bibinfo {author} {\bibfnamefont {R.~R.}\ \bibnamefont
  {da~Silva}}, \ and\ \bibinfo {author} {\bibfnamefont {S.}~\bibnamefont
  {Moehlecke}},\ }\href {\doibase 10.1103/PhysRevB.73.165128} {\bibfield
  {journal} {\bibinfo  {journal} {Phys. Rev. B}\ }\textbf {\bibinfo {volume}
  {73}},\ \bibinfo {pages} {165128} (\bibinfo {year} {2006})}\BibitemShut
  {NoStop}%
\bibitem [{\citenamefont {Wang}\ \emph {et~al.}(2014)\citenamefont {Wang},
  \citenamefont {Graf}, \citenamefont {Li}, \citenamefont {Wang},\ and\
  \citenamefont {Petrovic}}]{Wang2014}%
  \BibitemOpen
  \bibfield  {author} {\bibinfo {author} {\bibfnamefont {K.}~\bibnamefont
  {Wang}}, \bibinfo {author} {\bibfnamefont {D.}~\bibnamefont {Graf}}, \bibinfo
  {author} {\bibfnamefont {L.}~\bibnamefont {Li}}, \bibinfo {author}
  {\bibfnamefont {L.}~\bibnamefont {Wang}}, \ and\ \bibinfo {author}
  {\bibfnamefont {C.}~\bibnamefont {Petrovic}},\ }\href {\doibase
  10.1038/srep07328} {\bibfield  {journal} {\bibinfo  {journal} {Sci. Rep.}\
  }\textbf {\bibinfo {volume} {4}} (\bibinfo {year} {2014}),\
  10.1038/srep07328}\BibitemShut {NoStop}%
\bibitem [{\citenamefont {Zhao}\ \emph {et~al.}(2015)\citenamefont {Zhao},
  \citenamefont {Liu}, \citenamefont {Yan}, \citenamefont {An}, \citenamefont
  {Liu}, \citenamefont {Zhang}, \citenamefont {Wang}, \citenamefont {Liu},
  \citenamefont {Jiang}, \citenamefont {Li}, \citenamefont {Wang},
  \citenamefont {Li}, \citenamefont {Mandrus}, \citenamefont {Xie},
  \citenamefont {Pan},\ and\ \citenamefont {Wang}}]{Zhao2015}%
  \BibitemOpen
  \bibfield  {author} {\bibinfo {author} {\bibfnamefont {Y.}~\bibnamefont
  {Zhao}}, \bibinfo {author} {\bibfnamefont {H.}~\bibnamefont {Liu}}, \bibinfo
  {author} {\bibfnamefont {J.}~\bibnamefont {Yan}}, \bibinfo {author}
  {\bibfnamefont {W.}~\bibnamefont {An}}, \bibinfo {author} {\bibfnamefont
  {J.}~\bibnamefont {Liu}}, \bibinfo {author} {\bibfnamefont {X.}~\bibnamefont
  {Zhang}}, \bibinfo {author} {\bibfnamefont {H.}~\bibnamefont {Wang}},
  \bibinfo {author} {\bibfnamefont {Y.}~\bibnamefont {Liu}}, \bibinfo {author}
  {\bibfnamefont {H.}~\bibnamefont {Jiang}}, \bibinfo {author} {\bibfnamefont
  {Q.}~\bibnamefont {Li}}, \bibinfo {author} {\bibfnamefont {Y.}~\bibnamefont
  {Wang}}, \bibinfo {author} {\bibfnamefont {X.-Z.}\ \bibnamefont {Li}},
  \bibinfo {author} {\bibfnamefont {D.}~\bibnamefont {Mandrus}}, \bibinfo
  {author} {\bibfnamefont {X.~C.}\ \bibnamefont {Xie}}, \bibinfo {author}
  {\bibfnamefont {M.}~\bibnamefont {Pan}}, \ and\ \bibinfo {author}
  {\bibfnamefont {J.}~\bibnamefont {Wang}},\ }\href {\doibase
  10.1103/PhysRevB.92.041104} {\bibfield  {journal} {\bibinfo  {journal} {Phys.
  Rev. B}\ }\textbf {\bibinfo {volume} {92}},\ \bibinfo {pages} {041104}
  (\bibinfo {year} {2015})}\BibitemShut {NoStop}%
\bibitem [{\citenamefont {Xiang}\ \emph {et~al.}(2015)\citenamefont {Xiang},
  \citenamefont {Veldhorst}, \citenamefont {Dou},\ and\ \citenamefont
  {Wang}}]{Xiang2015}%
  \BibitemOpen
  \bibfield  {author} {\bibinfo {author} {\bibfnamefont {F.-X.}\ \bibnamefont
  {Xiang}}, \bibinfo {author} {\bibfnamefont {M.}~\bibnamefont {Veldhorst}},
  \bibinfo {author} {\bibfnamefont {S.-X.}\ \bibnamefont {Dou}}, \ and\
  \bibinfo {author} {\bibfnamefont {X.-L.}\ \bibnamefont {Wang}},\ }\href
  {\doibase 10.1209/0295-5075/112/37009} {\bibfield  {journal} {\bibinfo
  {journal} {{EPL} (Europhysics Letters)}\ }\textbf {\bibinfo {volume} {112}},\
  \bibinfo {pages} {37009} (\bibinfo {year} {2015})}\BibitemShut {NoStop}%
\bibitem [{\citenamefont {Xu}\ \emph {et~al.}(1997)\citenamefont {Xu},
  \citenamefont {Husmann}, \citenamefont {Rosenbaum}, \citenamefont {Saboungi},
  \citenamefont {Enderby},\ and\ \citenamefont {Littlewood}}]{Xu1997}%
  \BibitemOpen
  \bibfield  {author} {\bibinfo {author} {\bibfnamefont {R.}~\bibnamefont
  {Xu}}, \bibinfo {author} {\bibfnamefont {A.}~\bibnamefont {Husmann}},
  \bibinfo {author} {\bibfnamefont {T.}~\bibnamefont {Rosenbaum}}, \bibinfo
  {author} {\bibfnamefont {M.}~\bibnamefont {Saboungi}}, \bibinfo {author}
  {\bibfnamefont {J.}~\bibnamefont {Enderby}}, \ and\ \bibinfo {author}
  {\bibfnamefont {P.}~\bibnamefont {Littlewood}},\ }\href@noop {} {\bibfield
  {journal} {\bibinfo  {journal} {Nature}\ }\textbf {\bibinfo {volume} {390}},\
  \bibinfo {pages} {57} (\bibinfo {year} {1997})}\BibitemShut {NoStop}%
\bibitem [{\citenamefont {Lee}\ \emph {et~al.}(2002)\citenamefont {Lee},
  \citenamefont {Rosenbaum}, \citenamefont {Saboungi},\ and\ \citenamefont
  {Schnyders}}]{Lee2002}%
  \BibitemOpen
  \bibfield  {author} {\bibinfo {author} {\bibfnamefont {M.}~\bibnamefont
  {Lee}}, \bibinfo {author} {\bibfnamefont {T.~F.}\ \bibnamefont {Rosenbaum}},
  \bibinfo {author} {\bibfnamefont {M.-L.}\ \bibnamefont {Saboungi}}, \ and\
  \bibinfo {author} {\bibfnamefont {H.~S.}\ \bibnamefont {Schnyders}},\ }\href
  {\doibase 10.1103/PhysRevLett.88.066602} {\bibfield  {journal} {\bibinfo
  {journal} {Phys. Rev. Lett.}\ }\textbf {\bibinfo {volume} {88}},\ \bibinfo
  {pages} {066602} (\bibinfo {year} {2002})}\BibitemShut {NoStop}%
\bibitem [{\citenamefont {Singh}\ \emph {et~al.}(2012)\citenamefont {Singh},
  \citenamefont {Wang}, \citenamefont {Chen}, \citenamefont {Ariando},\ and\
  \citenamefont {Wee}}]{Singh2012}%
  \BibitemOpen
  \bibfield  {author} {\bibinfo {author} {\bibfnamefont {R.~S.}\ \bibnamefont
  {Singh}}, \bibinfo {author} {\bibfnamefont {X.}~\bibnamefont {Wang}},
  \bibinfo {author} {\bibfnamefont {W.}~\bibnamefont {Chen}}, \bibinfo {author}
  {\bibnamefont {Ariando}}, \ and\ \bibinfo {author} {\bibfnamefont {A.~T.~S.}\
  \bibnamefont {Wee}},\ }\href {\doibase 10.1063/1.4765656} {\bibfield
  {journal} {\bibinfo  {journal} {Applied Physics Letters}\ }\textbf {\bibinfo
  {volume} {101}},\ \bibinfo {pages} {183105} (\bibinfo {year}
  {2012})}\BibitemShut {NoStop}%
\bibitem [{\citenamefont {Qu}\ \emph {et~al.}(2010)\citenamefont {Qu},
  \citenamefont {Hor}, \citenamefont {Xiong}, \citenamefont {Cava},\ and\
  \citenamefont {Ong}}]{Qu2010}%
  \BibitemOpen
  \bibfield  {author} {\bibinfo {author} {\bibfnamefont {D.-X.}\ \bibnamefont
  {Qu}}, \bibinfo {author} {\bibfnamefont {Y.~S.}\ \bibnamefont {Hor}},
  \bibinfo {author} {\bibfnamefont {J.}~\bibnamefont {Xiong}}, \bibinfo
  {author} {\bibfnamefont {R.~J.}\ \bibnamefont {Cava}}, \ and\ \bibinfo
  {author} {\bibfnamefont {N.~P.}\ \bibnamefont {Ong}},\ }\href {\doibase
  10.1126/science.1189792} {\bibfield  {journal} {\bibinfo  {journal}
  {Science}\ }\textbf {\bibinfo {volume} {329}},\ \bibinfo {pages} {821}
  (\bibinfo {year} {2010})}\BibitemShut {NoStop}%
\bibitem [{\citenamefont {Wang}\ \emph {et~al.}(2012)\citenamefont {Wang},
  \citenamefont {Du}, \citenamefont {Dou},\ and\ \citenamefont
  {Zhang}}]{Wang2012b}%
  \BibitemOpen
  \bibfield  {author} {\bibinfo {author} {\bibfnamefont {X.}~\bibnamefont
  {Wang}}, \bibinfo {author} {\bibfnamefont {Y.}~\bibnamefont {Du}}, \bibinfo
  {author} {\bibfnamefont {S.}~\bibnamefont {Dou}}, \ and\ \bibinfo {author}
  {\bibfnamefont {C.}~\bibnamefont {Zhang}},\ }\href {\doibase
  10.1103/PhysRevLett.108.266806} {\bibfield  {journal} {\bibinfo  {journal}
  {Phys. Rev. Lett.}\ }\textbf {\bibinfo {volume} {108}},\ \bibinfo {pages}
  {266806} (\bibinfo {year} {2012})}\BibitemShut {NoStop}%
\bibitem [{\citenamefont {Liang}\ \emph {et~al.}(2014)\citenamefont {Liang},
  \citenamefont {Gibson}, \citenamefont {Ali}, \citenamefont {Liu},
  \citenamefont {Cava},\ and\ \citenamefont {Ong}}]{Liang2014}%
  \BibitemOpen
  \bibfield  {author} {\bibinfo {author} {\bibfnamefont {T.}~\bibnamefont
  {Liang}}, \bibinfo {author} {\bibfnamefont {Q.}~\bibnamefont {Gibson}},
  \bibinfo {author} {\bibfnamefont {M.~N.}\ \bibnamefont {Ali}}, \bibinfo
  {author} {\bibfnamefont {M.}~\bibnamefont {Liu}}, \bibinfo {author}
  {\bibfnamefont {R.~J.}\ \bibnamefont {Cava}}, \ and\ \bibinfo {author}
  {\bibfnamefont {N.~P.}\ \bibnamefont {Ong}},\ }\href {\doibase
  10.1038/nmat4143} {\bibfield  {journal} {\bibinfo  {journal} {Nature
  Materials}\ }\textbf {\bibinfo {volume} {14}},\ \bibinfo {pages} {280}
  (\bibinfo {year} {2014})}\BibitemShut {NoStop}%
\bibitem [{\citenamefont {Feng}\ \emph {et~al.}(2015)\citenamefont {Feng},
  \citenamefont {Pang}, \citenamefont {Wu}, \citenamefont {Wang}, \citenamefont
  {Weng}, \citenamefont {Li}, \citenamefont {Dai}, \citenamefont {Fang},
  \citenamefont {Shi},\ and\ \citenamefont {Lu}}]{Feng2015}%
  \BibitemOpen
  \bibfield  {author} {\bibinfo {author} {\bibfnamefont {J.}~\bibnamefont
  {Feng}}, \bibinfo {author} {\bibfnamefont {Y.}~\bibnamefont {Pang}}, \bibinfo
  {author} {\bibfnamefont {D.}~\bibnamefont {Wu}}, \bibinfo {author}
  {\bibfnamefont {Z.}~\bibnamefont {Wang}}, \bibinfo {author} {\bibfnamefont
  {H.}~\bibnamefont {Weng}}, \bibinfo {author} {\bibfnamefont {J.}~\bibnamefont
  {Li}}, \bibinfo {author} {\bibfnamefont {X.}~\bibnamefont {Dai}}, \bibinfo
  {author} {\bibfnamefont {Z.}~\bibnamefont {Fang}}, \bibinfo {author}
  {\bibfnamefont {Y.}~\bibnamefont {Shi}}, \ and\ \bibinfo {author}
  {\bibfnamefont {L.}~\bibnamefont {Lu}},\ }\href {\doibase
  10.1103/PhysRevB.92.081306} {\bibfield  {journal} {\bibinfo  {journal} {Phys.
  Rev. B}\ }\textbf {\bibinfo {volume} {92}},\ \bibinfo {pages} {081306}
  (\bibinfo {year} {2015})}\BibitemShut {NoStop}%
\bibitem [{\citenamefont {Kushwaha}\ \emph {et~al.}(2015)\citenamefont
  {Kushwaha}, \citenamefont {Krizan}, \citenamefont {Feldman}, \citenamefont
  {Gyenis}, \citenamefont {Randeria}, \citenamefont {Xiong}, \citenamefont
  {Xu}, \citenamefont {Alidoust}, \citenamefont {Belopolski}, \citenamefont
  {Liang}, \citenamefont {Hasan}, \citenamefont {Ong}, \citenamefont
  {Yazdani},\ and\ \citenamefont {Cava}}]{Kushwaha2015}%
  \BibitemOpen
  \bibfield  {author} {\bibinfo {author} {\bibfnamefont {S.~K.}\ \bibnamefont
  {Kushwaha}}, \bibinfo {author} {\bibfnamefont {J.~W.}\ \bibnamefont
  {Krizan}}, \bibinfo {author} {\bibfnamefont {B.~E.}\ \bibnamefont {Feldman}},
  \bibinfo {author} {\bibfnamefont {A.}~\bibnamefont {Gyenis}}, \bibinfo
  {author} {\bibfnamefont {M.~T.}\ \bibnamefont {Randeria}}, \bibinfo {author}
  {\bibfnamefont {J.}~\bibnamefont {Xiong}}, \bibinfo {author} {\bibfnamefont
  {S.-Y.}\ \bibnamefont {Xu}}, \bibinfo {author} {\bibfnamefont
  {N.}~\bibnamefont {Alidoust}}, \bibinfo {author} {\bibfnamefont
  {I.}~\bibnamefont {Belopolski}}, \bibinfo {author} {\bibfnamefont
  {T.}~\bibnamefont {Liang}}, \bibinfo {author} {\bibfnamefont {M.~Z.}\
  \bibnamefont {Hasan}}, \bibinfo {author} {\bibfnamefont {N.~P.}\ \bibnamefont
  {Ong}}, \bibinfo {author} {\bibfnamefont {A.}~\bibnamefont {Yazdani}}, \ and\
  \bibinfo {author} {\bibfnamefont {R.~J.}\ \bibnamefont {Cava}},\ }\href
  {\doibase 10.1063/1.4908158} {\bibfield  {journal} {\bibinfo  {journal}
  {{APL} Materials}\ }\textbf {\bibinfo {volume} {3}},\ \bibinfo {pages}
  {041504} (\bibinfo {year} {2015})}\BibitemShut {NoStop}%
\bibitem [{\citenamefont {Huang}\ \emph {et~al.}(2015)\citenamefont {Huang},
  \citenamefont {Zhao}, \citenamefont {Long}, \citenamefont {Wang},
  \citenamefont {Chen}, \citenamefont {Yang}, \citenamefont {Liang},
  \citenamefont {Xue}, \citenamefont {Weng}, \citenamefont {Fang},
  \citenamefont {Dai},\ and\ \citenamefont {Chen}}]{Huang2015}%
  \BibitemOpen
  \bibfield  {author} {\bibinfo {author} {\bibfnamefont {X.}~\bibnamefont
  {Huang}}, \bibinfo {author} {\bibfnamefont {L.}~\bibnamefont {Zhao}},
  \bibinfo {author} {\bibfnamefont {Y.}~\bibnamefont {Long}}, \bibinfo {author}
  {\bibfnamefont {P.}~\bibnamefont {Wang}}, \bibinfo {author} {\bibfnamefont
  {D.}~\bibnamefont {Chen}}, \bibinfo {author} {\bibfnamefont {Z.}~\bibnamefont
  {Yang}}, \bibinfo {author} {\bibfnamefont {H.}~\bibnamefont {Liang}},
  \bibinfo {author} {\bibfnamefont {M.}~\bibnamefont {Xue}}, \bibinfo {author}
  {\bibfnamefont {H.}~\bibnamefont {Weng}}, \bibinfo {author} {\bibfnamefont
  {Z.}~\bibnamefont {Fang}}, \bibinfo {author} {\bibfnamefont {X.}~\bibnamefont
  {Dai}}, \ and\ \bibinfo {author} {\bibfnamefont {G.}~\bibnamefont {Chen}},\
  }\href {\doibase 10.1103/PhysRevX.5.031023} {\bibfield  {journal} {\bibinfo
  {journal} {Phys. Rev. X}\ }\textbf {\bibinfo {volume} {5}},\ \bibinfo {pages}
  {031023} (\bibinfo {year} {2015})}\BibitemShut {NoStop}%
\bibitem [{\citenamefont {{Zhang}}\ \emph {et~al.}(2015)\citenamefont
  {{Zhang}}, \citenamefont {{Yuan}}, \citenamefont {{Xu}}, \citenamefont
  {{Lin}}, \citenamefont {{Tong}}, \citenamefont {{Zahid Hasan}}, \citenamefont
  {{Wang}}, \citenamefont {{Zhang}},\ and\ \citenamefont {{Jia}}}]{Zhang2015a}%
  \BibitemOpen
  \bibfield  {author} {\bibinfo {author} {\bibfnamefont {C.}~\bibnamefont
  {{Zhang}}}, \bibinfo {author} {\bibfnamefont {Z.}~\bibnamefont {{Yuan}}},
  \bibinfo {author} {\bibfnamefont {S.}~\bibnamefont {{Xu}}}, \bibinfo {author}
  {\bibfnamefont {Z.}~\bibnamefont {{Lin}}}, \bibinfo {author} {\bibfnamefont
  {B.}~\bibnamefont {{Tong}}}, \bibinfo {author} {\bibfnamefont
  {M.}~\bibnamefont {{Zahid Hasan}}}, \bibinfo {author} {\bibfnamefont
  {J.}~\bibnamefont {{Wang}}}, \bibinfo {author} {\bibfnamefont
  {C.}~\bibnamefont {{Zhang}}}, \ and\ \bibinfo {author} {\bibfnamefont
  {S.}~\bibnamefont {{Jia}}},\ }\href@noop {} {\bibfield  {journal} {\bibinfo
  {journal} {ArXiv e-prints}\ } (\bibinfo {year} {2015})},\ \Eprint
  {http://arxiv.org/abs/1502.00251} {arXiv:1502.00251 [cond-mat.mtrl-sci]}
  \BibitemShut {NoStop}%
\bibitem [{\citenamefont {Ghimire}\ \emph {et~al.}(2015)\citenamefont
  {Ghimire}, \citenamefont {Luo}, \citenamefont {Neupane}, \citenamefont
  {Williams}, \citenamefont {Bauer},\ and\ \citenamefont
  {Ronning}}]{Ghimire2015}%
  \BibitemOpen
  \bibfield  {author} {\bibinfo {author} {\bibfnamefont {N.~J.}\ \bibnamefont
  {Ghimire}}, \bibinfo {author} {\bibfnamefont {Y.}~\bibnamefont {Luo}},
  \bibinfo {author} {\bibfnamefont {M.}~\bibnamefont {Neupane}}, \bibinfo
  {author} {\bibfnamefont {D.~J.}\ \bibnamefont {Williams}}, \bibinfo {author}
  {\bibfnamefont {E.~D.}\ \bibnamefont {Bauer}}, \ and\ \bibinfo {author}
  {\bibfnamefont {F.}~\bibnamefont {Ronning}},\ }\href {\doibase
  10.1088/0953-8984/27/15/152201} {\bibfield  {journal} {\bibinfo  {journal}
  {Journal of Physics: Condensed Matter}\ }\textbf {\bibinfo {volume} {27}},\
  \bibinfo {pages} {152201} (\bibinfo {year} {2015})}\BibitemShut {NoStop}%
\bibitem [{\citenamefont {Shekhar}\ \emph {et~al.}(2015)\citenamefont
  {Shekhar}, \citenamefont {Nayak}, \citenamefont {Sun}, \citenamefont
  {Schmidt}, \citenamefont {Nicklas}, \citenamefont {Leermakers}, \citenamefont
  {Zeitler}, \citenamefont {Skourski}, \citenamefont {Wosnitza}, \citenamefont
  {Liu}, \citenamefont {Chen}, \citenamefont {Schnelle}, \citenamefont
  {Borrmann}, \citenamefont {Grin}, \citenamefont {Felser},\ and\ \citenamefont
  {Yan}}]{Shekhar2015}%
  \BibitemOpen
  \bibfield  {author} {\bibinfo {author} {\bibfnamefont {C.}~\bibnamefont
  {Shekhar}}, \bibinfo {author} {\bibfnamefont {A.~K.}\ \bibnamefont {Nayak}},
  \bibinfo {author} {\bibfnamefont {Y.}~\bibnamefont {Sun}}, \bibinfo {author}
  {\bibfnamefont {M.}~\bibnamefont {Schmidt}}, \bibinfo {author} {\bibfnamefont
  {M.}~\bibnamefont {Nicklas}}, \bibinfo {author} {\bibfnamefont
  {I.}~\bibnamefont {Leermakers}}, \bibinfo {author} {\bibfnamefont
  {U.}~\bibnamefont {Zeitler}}, \bibinfo {author} {\bibfnamefont
  {Y.}~\bibnamefont {Skourski}}, \bibinfo {author} {\bibfnamefont
  {J.}~\bibnamefont {Wosnitza}}, \bibinfo {author} {\bibfnamefont
  {Z.}~\bibnamefont {Liu}}, \bibinfo {author} {\bibfnamefont {Y.}~\bibnamefont
  {Chen}}, \bibinfo {author} {\bibfnamefont {W.}~\bibnamefont {Schnelle}},
  \bibinfo {author} {\bibfnamefont {H.}~\bibnamefont {Borrmann}}, \bibinfo
  {author} {\bibfnamefont {Y.}~\bibnamefont {Grin}}, \bibinfo {author}
  {\bibfnamefont {C.}~\bibnamefont {Felser}}, \ and\ \bibinfo {author}
  {\bibfnamefont {B.}~\bibnamefont {Yan}},\ }\href {\doibase 10.1038/nphys3372}
  {\bibfield  {journal} {\bibinfo  {journal} {Nature Physics}\ }\textbf
  {\bibinfo {volume} {11}},\ \bibinfo {pages} {645} (\bibinfo {year}
  {2015})}\BibitemShut {NoStop}%
\bibitem [{\citenamefont {Wang}\ \emph
  {et~al.}(2016{\natexlab{a}})\citenamefont {Wang}, \citenamefont {Zheng},
  \citenamefont {Shen}, \citenamefont {Lu}, \citenamefont {Fang}, \citenamefont
  {Sheng}, \citenamefont {Zhou}, \citenamefont {Yang}, \citenamefont {Li},
  \citenamefont {Feng},\ and\ \citenamefont {Xu}}]{Wang2016a}%
  \BibitemOpen
  \bibfield  {author} {\bibinfo {author} {\bibfnamefont {Z.}~\bibnamefont
  {Wang}}, \bibinfo {author} {\bibfnamefont {Y.}~\bibnamefont {Zheng}},
  \bibinfo {author} {\bibfnamefont {Z.}~\bibnamefont {Shen}}, \bibinfo {author}
  {\bibfnamefont {Y.}~\bibnamefont {Lu}}, \bibinfo {author} {\bibfnamefont
  {H.}~\bibnamefont {Fang}}, \bibinfo {author} {\bibfnamefont {F.}~\bibnamefont
  {Sheng}}, \bibinfo {author} {\bibfnamefont {Y.}~\bibnamefont {Zhou}},
  \bibinfo {author} {\bibfnamefont {X.}~\bibnamefont {Yang}}, \bibinfo {author}
  {\bibfnamefont {Y.}~\bibnamefont {Li}}, \bibinfo {author} {\bibfnamefont
  {C.}~\bibnamefont {Feng}}, \ and\ \bibinfo {author} {\bibfnamefont {Z.-A.}\
  \bibnamefont {Xu}},\ }\href {\doibase 10.1103/PhysRevB.93.121112} {\bibfield
  {journal} {\bibinfo  {journal} {Phys. Rev. B}\ }\textbf {\bibinfo {volume}
  {93}},\ \bibinfo {pages} {121112} (\bibinfo {year}
  {2016}{\natexlab{a}})}\BibitemShut {NoStop}%
\bibitem [{\citenamefont {Abrikosov}(1998)}]{Abrikosov1998}%
  \BibitemOpen
  \bibfield  {author} {\bibinfo {author} {\bibfnamefont {A.~A.}\ \bibnamefont
  {Abrikosov}},\ }\href {\doibase 10.1103/physrevb.58.2788} {\bibfield
  {journal} {\bibinfo  {journal} {Physical Review B}\ }\textbf {\bibinfo
  {volume} {58}},\ \bibinfo {pages} {2788} (\bibinfo {year}
  {1998})}\BibitemShut {NoStop}%
\bibitem [{\citenamefont {Abrikosov}(2003)}]{Abrikosov2003}%
  \BibitemOpen
  \bibfield  {author} {\bibinfo {author} {\bibfnamefont {A.~A.}\ \bibnamefont
  {Abrikosov}},\ }\href {\doibase 10.1088/0305-4470/36/35/301} {\bibfield
  {journal} {\bibinfo  {journal} {J. Phys. A: Math. Gen.}\ }\textbf {\bibinfo
  {volume} {36}},\ \bibinfo {pages} {9119} (\bibinfo {year}
  {2003})}\BibitemShut {NoStop}%
\bibitem [{\citenamefont {Zeng}\ \emph {et~al.}(2016)\citenamefont {Zeng},
  \citenamefont {Lou}, \citenamefont {Wu}, \citenamefont {Xu}, \citenamefont
  {Guo}, \citenamefont {Kong}, \citenamefont {Zhong}, \citenamefont {Ma},
  \citenamefont {Fu}, \citenamefont {Richard}, \citenamefont {Wang},
  \citenamefont {Liu}, \citenamefont {Lu}, \citenamefont {Huang}, \citenamefont
  {Fang}, \citenamefont {Sun}, \citenamefont {Wang}, \citenamefont {Wang},
  \citenamefont {Shi}, \citenamefont {Weng}, \citenamefont {Lei}, \citenamefont
  {Liu}, \citenamefont {Wang}, \citenamefont {Qian}, \citenamefont {Luo},\ and\
  \citenamefont {Ding}}]{Zeng2016}%
  \BibitemOpen
  \bibfield  {author} {\bibinfo {author} {\bibfnamefont {L.-K.}\ \bibnamefont
  {Zeng}}, \bibinfo {author} {\bibfnamefont {R.}~\bibnamefont {Lou}}, \bibinfo
  {author} {\bibfnamefont {D.-S.}\ \bibnamefont {Wu}}, \bibinfo {author}
  {\bibfnamefont {Q.~N.}\ \bibnamefont {Xu}}, \bibinfo {author} {\bibfnamefont
  {P.-J.}\ \bibnamefont {Guo}}, \bibinfo {author} {\bibfnamefont {L.-Y.}\
  \bibnamefont {Kong}}, \bibinfo {author} {\bibfnamefont {Y.-G.}\ \bibnamefont
  {Zhong}}, \bibinfo {author} {\bibfnamefont {J.-Z.}\ \bibnamefont {Ma}},
  \bibinfo {author} {\bibfnamefont {B.-B.}\ \bibnamefont {Fu}}, \bibinfo
  {author} {\bibfnamefont {P.}~\bibnamefont {Richard}}, \bibinfo {author}
  {\bibfnamefont {P.}~\bibnamefont {Wang}}, \bibinfo {author} {\bibfnamefont
  {G.~T.}\ \bibnamefont {Liu}}, \bibinfo {author} {\bibfnamefont
  {L.}~\bibnamefont {Lu}}, \bibinfo {author} {\bibfnamefont {Y.-B.}\
  \bibnamefont {Huang}}, \bibinfo {author} {\bibfnamefont {C.}~\bibnamefont
  {Fang}}, \bibinfo {author} {\bibfnamefont {S.-S.}\ \bibnamefont {Sun}},
  \bibinfo {author} {\bibfnamefont {Q.}~\bibnamefont {Wang}}, \bibinfo {author}
  {\bibfnamefont {L.}~\bibnamefont {Wang}}, \bibinfo {author} {\bibfnamefont
  {Y.-G.}\ \bibnamefont {Shi}}, \bibinfo {author} {\bibfnamefont {H.~M.}\
  \bibnamefont {Weng}}, \bibinfo {author} {\bibfnamefont {H.-C.}\ \bibnamefont
  {Lei}}, \bibinfo {author} {\bibfnamefont {K.}~\bibnamefont {Liu}}, \bibinfo
  {author} {\bibfnamefont {S.-C.}\ \bibnamefont {Wang}}, \bibinfo {author}
  {\bibfnamefont {T.}~\bibnamefont {Qian}}, \bibinfo {author} {\bibfnamefont
  {J.-L.}\ \bibnamefont {Luo}}, \ and\ \bibinfo {author} {\bibfnamefont
  {H.}~\bibnamefont {Ding}},\ }\href {\doibase 10.1103/PhysRevLett.117.127204}
  {\bibfield  {journal} {\bibinfo  {journal} {Phys. Rev. Lett.}\ }\textbf
  {\bibinfo {volume} {117}},\ \bibinfo {pages} {127204} (\bibinfo {year}
  {2016})}\BibitemShut {NoStop}%
\bibitem [{\citenamefont {Lv}\ \emph {et~al.}(2016)\citenamefont {Lv},
  \citenamefont {Zhang}, \citenamefont {Li}, \citenamefont {Yao}, \citenamefont
  {Chen}, \citenamefont {Zhou}, \citenamefont {Zhang}, \citenamefont {Lu},\
  and\ \citenamefont {Chen}}]{Lv2016}%
  \BibitemOpen
  \bibfield  {author} {\bibinfo {author} {\bibfnamefont {Y.-Y.}\ \bibnamefont
  {Lv}}, \bibinfo {author} {\bibfnamefont {B.-B.}\ \bibnamefont {Zhang}},
  \bibinfo {author} {\bibfnamefont {X.}~\bibnamefont {Li}}, \bibinfo {author}
  {\bibfnamefont {S.-H.}\ \bibnamefont {Yao}}, \bibinfo {author} {\bibfnamefont
  {Y.~B.}\ \bibnamefont {Chen}}, \bibinfo {author} {\bibfnamefont
  {J.}~\bibnamefont {Zhou}}, \bibinfo {author} {\bibfnamefont {S.-T.}\
  \bibnamefont {Zhang}}, \bibinfo {author} {\bibfnamefont {M.-H.}\ \bibnamefont
  {Lu}}, \ and\ \bibinfo {author} {\bibfnamefont {Y.-F.}\ \bibnamefont
  {Chen}},\ }\href {\doibase 10.1063/1.4953772} {\bibfield  {journal} {\bibinfo
   {journal} {Applied Physics Letters}\ }\textbf {\bibinfo {volume} {108}},\
  \bibinfo {pages} {244101} (\bibinfo {year} {2016})}\BibitemShut {NoStop}%
\bibitem [{\citenamefont {Zandt}\ \emph {et~al.}(2007)\citenamefont {Zandt},
  \citenamefont {Dwelk}, \citenamefont {Janowitz},\ and\ \citenamefont
  {Manzke}}]{Zandt2007}%
  \BibitemOpen
  \bibfield  {author} {\bibinfo {author} {\bibfnamefont {T.}~\bibnamefont
  {Zandt}}, \bibinfo {author} {\bibfnamefont {H.}~\bibnamefont {Dwelk}},
  \bibinfo {author} {\bibfnamefont {C.}~\bibnamefont {Janowitz}}, \ and\
  \bibinfo {author} {\bibfnamefont {R.}~\bibnamefont {Manzke}},\ }\href
  {\doibase http://dx.doi.org/10.1016/j.jallcom.2006.09.157} {\bibfield
  {journal} {\bibinfo  {journal} {J. Alloys Compd.}\ }\textbf {\bibinfo
  {volume} {442}},\ \bibinfo {pages} {216 } (\bibinfo {year}
  {2007})}\BibitemShut {NoStop}%
\bibitem [{\citenamefont {S.~Thirupathaiah}(2017)}]{Thirupathaiah2017}%
  \BibitemOpen
  \bibfield  {author} {\bibinfo {author} {\bibfnamefont {R.~J. B. P. J. S. M.
  P. K. D. P. K. S. I. V. N. C. P. M. S. R. A. R. D. D.~S.}\ \bibnamefont
  {S.~Thirupathaiah}},\ }\href@noop {} {\bibfield  {journal} {\bibinfo
  {journal} {arXiv:1705.07217}\ } (\bibinfo {year} {2017})}\BibitemShut
  {NoStop}%
\bibitem [{\citenamefont {Pletikosi\ifmmode~\acute{c}\else \'{c}\fi{}}\ \emph
  {et~al.}(2014)\citenamefont {Pletikosi\ifmmode~\acute{c}\else \'{c}\fi{}},
  \citenamefont {Ali}, \citenamefont {Fedorov}, \citenamefont {Cava},\ and\
  \citenamefont {Valla}}]{Pletikosic2014}%
  \BibitemOpen
  \bibfield  {author} {\bibinfo {author} {\bibfnamefont {I.}~\bibnamefont
  {Pletikosi\ifmmode~\acute{c}\else \'{c}\fi{}}}, \bibinfo {author}
  {\bibfnamefont {M.~N.}\ \bibnamefont {Ali}}, \bibinfo {author} {\bibfnamefont
  {A.~V.}\ \bibnamefont {Fedorov}}, \bibinfo {author} {\bibfnamefont {R.~J.}\
  \bibnamefont {Cava}}, \ and\ \bibinfo {author} {\bibfnamefont
  {T.}~\bibnamefont {Valla}},\ }\href {\doibase 10.1103/PhysRevLett.113.216601}
  {\bibfield  {journal} {\bibinfo  {journal} {Phys. Rev. Lett.}\ }\textbf
  {\bibinfo {volume} {113}},\ \bibinfo {pages} {216601} (\bibinfo {year}
  {2014})}\BibitemShut {NoStop}%
\bibitem [{\citenamefont {Wu}\ \emph {et~al.}(2015)\citenamefont {Wu},
  \citenamefont {Jo}, \citenamefont {Ochi}, \citenamefont {Huang},
  \citenamefont {Mou}, \citenamefont {Bud'ko}, \citenamefont {Canfield},
  \citenamefont {Trivedi}, \citenamefont {Arita},\ and\ \citenamefont
  {Kaminski}}]{Wu2015}%
  \BibitemOpen
  \bibfield  {author} {\bibinfo {author} {\bibfnamefont {Y.}~\bibnamefont
  {Wu}}, \bibinfo {author} {\bibfnamefont {N.~H.}\ \bibnamefont {Jo}}, \bibinfo
  {author} {\bibfnamefont {M.}~\bibnamefont {Ochi}}, \bibinfo {author}
  {\bibfnamefont {L.}~\bibnamefont {Huang}}, \bibinfo {author} {\bibfnamefont
  {D.}~\bibnamefont {Mou}}, \bibinfo {author} {\bibfnamefont {S.~L.}\
  \bibnamefont {Bud'ko}}, \bibinfo {author} {\bibfnamefont {P.~C.}\
  \bibnamefont {Canfield}}, \bibinfo {author} {\bibfnamefont {N.}~\bibnamefont
  {Trivedi}}, \bibinfo {author} {\bibfnamefont {R.}~\bibnamefont {Arita}}, \
  and\ \bibinfo {author} {\bibfnamefont {A.}~\bibnamefont {Kaminski}},\ }\href
  {\doibase 10.1103/PhysRevLett.115.166602} {\bibfield  {journal} {\bibinfo
  {journal} {Phys. Rev. Lett.}\ }\textbf {\bibinfo {volume} {115}},\ \bibinfo
  {pages} {166602} (\bibinfo {year} {2015})}\BibitemShut {NoStop}%
\bibitem [{\citenamefont {Kabashima}(1966)}]{Kabashima1966}%
  \BibitemOpen
  \bibfield  {author} {\bibinfo {author} {\bibfnamefont {S.}~\bibnamefont
  {Kabashima}},\ }\href {\doibase 10.1143/jpsj.21.945} {\bibfield  {journal}
  {\bibinfo  {journal} {J. Phys. Soc. Jpn.}\ }\textbf {\bibinfo {volume}
  {21}},\ \bibinfo {pages} {945} (\bibinfo {year} {1966})}\BibitemShut
  {NoStop}%
\bibitem [{\citenamefont {Liu}\ \emph {et~al.}(2017)\citenamefont {Liu},
  \citenamefont {Liu}, \citenamefont {Zhou},\ and\ \citenamefont
  {Wan}}]{Liu2017}%
  \BibitemOpen
  \bibfield  {author} {\bibinfo {author} {\bibfnamefont {G.}~\bibnamefont
  {Liu}}, \bibinfo {author} {\bibfnamefont {H.}~\bibnamefont {Liu}}, \bibinfo
  {author} {\bibfnamefont {J.}~\bibnamefont {Zhou}}, \ and\ \bibinfo {author}
  {\bibfnamefont {X.}~\bibnamefont {Wan}},\ }\href {\doibase 10.1063/1.4974946}
  {\bibfield  {journal} {\bibinfo  {journal} {J. Appl. Phys.}\ }\textbf
  {\bibinfo {volume} {121}},\ \bibinfo {pages} {045104} (\bibinfo {year}
  {2017})}\BibitemShut {NoStop}%
\bibitem [{\citenamefont {Jiang}\ \emph {et~al.}(2015)\citenamefont {Jiang},
  \citenamefont {Tang}, \citenamefont {Pan}, \citenamefont {Liu}, \citenamefont
  {Niu}, \citenamefont {Wang}, \citenamefont {Xu}, \citenamefont {Yang},
  \citenamefont {Xie}, \citenamefont {Song}, \citenamefont {Dudin},
  \citenamefont {Kim}, \citenamefont {Hoesch}, \citenamefont {Das},
  \citenamefont {Vobornik}, \citenamefont {Wan},\ and\ \citenamefont
  {Feng}}]{Jiang2015}%
  \BibitemOpen
  \bibfield  {author} {\bibinfo {author} {\bibfnamefont {J.}~\bibnamefont
  {Jiang}}, \bibinfo {author} {\bibfnamefont {F.}~\bibnamefont {Tang}},
  \bibinfo {author} {\bibfnamefont {X.~C.}\ \bibnamefont {Pan}}, \bibinfo
  {author} {\bibfnamefont {H.~M.}\ \bibnamefont {Liu}}, \bibinfo {author}
  {\bibfnamefont {X.~H.}\ \bibnamefont {Niu}}, \bibinfo {author} {\bibfnamefont
  {Y.~X.}\ \bibnamefont {Wang}}, \bibinfo {author} {\bibfnamefont {D.~F.}\
  \bibnamefont {Xu}}, \bibinfo {author} {\bibfnamefont {H.~F.}\ \bibnamefont
  {Yang}}, \bibinfo {author} {\bibfnamefont {B.~P.}\ \bibnamefont {Xie}},
  \bibinfo {author} {\bibfnamefont {F.~Q.}\ \bibnamefont {Song}}, \bibinfo
  {author} {\bibfnamefont {P.}~\bibnamefont {Dudin}}, \bibinfo {author}
  {\bibfnamefont {T.~K.}\ \bibnamefont {Kim}}, \bibinfo {author} {\bibfnamefont
  {M.}~\bibnamefont {Hoesch}}, \bibinfo {author} {\bibfnamefont {P.~K.}\
  \bibnamefont {Das}}, \bibinfo {author} {\bibfnamefont {I.}~\bibnamefont
  {Vobornik}}, \bibinfo {author} {\bibfnamefont {X.~G.}\ \bibnamefont {Wan}}, \
  and\ \bibinfo {author} {\bibfnamefont {D.~L.}\ \bibnamefont {Feng}},\ }\href
  {\doibase 10.1103/PhysRevLett.115.166601} {\bibfield  {journal} {\bibinfo
  {journal} {Phys. Rev. Lett.}\ }\textbf {\bibinfo {volume} {115}},\ \bibinfo
  {pages} {166601} (\bibinfo {year} {2015})}\BibitemShut {NoStop}%
\bibitem [{\citenamefont {Bruno}\ \emph {et~al.}(2016)\citenamefont {Bruno},
  \citenamefont {Tamai}, \citenamefont {Wu}, \citenamefont {Cucchi},
  \citenamefont {Barreteau}, \citenamefont {de~la Torre}, \citenamefont
  {McKeown~Walker}, \citenamefont {Ricc\`o}, \citenamefont {Wang},
  \citenamefont {Kim}, \citenamefont {Hoesch}, \citenamefont {Shi},
  \citenamefont {Plumb}, \citenamefont {Giannini}, \citenamefont {Soluyanov},\
  and\ \citenamefont {Baumberger}}]{Bruno2016}%
  \BibitemOpen
  \bibfield  {author} {\bibinfo {author} {\bibfnamefont {F.~Y.}\ \bibnamefont
  {Bruno}}, \bibinfo {author} {\bibfnamefont {A.}~\bibnamefont {Tamai}},
  \bibinfo {author} {\bibfnamefont {Q.~S.}\ \bibnamefont {Wu}}, \bibinfo
  {author} {\bibfnamefont {I.}~\bibnamefont {Cucchi}}, \bibinfo {author}
  {\bibfnamefont {C.}~\bibnamefont {Barreteau}}, \bibinfo {author}
  {\bibfnamefont {A.}~\bibnamefont {de~la Torre}}, \bibinfo {author}
  {\bibfnamefont {S.}~\bibnamefont {McKeown~Walker}}, \bibinfo {author}
  {\bibfnamefont {S.}~\bibnamefont {Ricc\`o}}, \bibinfo {author} {\bibfnamefont
  {Z.}~\bibnamefont {Wang}}, \bibinfo {author} {\bibfnamefont {T.~K.}\
  \bibnamefont {Kim}}, \bibinfo {author} {\bibfnamefont {M.}~\bibnamefont
  {Hoesch}}, \bibinfo {author} {\bibfnamefont {M.}~\bibnamefont {Shi}},
  \bibinfo {author} {\bibfnamefont {N.~C.}\ \bibnamefont {Plumb}}, \bibinfo
  {author} {\bibfnamefont {E.}~\bibnamefont {Giannini}}, \bibinfo {author}
  {\bibfnamefont {A.~A.}\ \bibnamefont {Soluyanov}}, \ and\ \bibinfo {author}
  {\bibfnamefont {F.}~\bibnamefont {Baumberger}},\ }\href {\doibase
  10.1103/PhysRevB.94.121112} {\bibfield  {journal} {\bibinfo  {journal} {Phys.
  Rev. B}\ }\textbf {\bibinfo {volume} {94}},\ \bibinfo {pages} {121112 (R)}
  (\bibinfo {year} {2016})}\BibitemShut {NoStop}%
\bibitem [{\citenamefont {Wu}\ \emph {et~al.}(2016)\citenamefont {Wu},
  \citenamefont {Mou}, \citenamefont {Jo}, \citenamefont {Sun}, \citenamefont
  {Huang}, \citenamefont {Bud'ko}, \citenamefont {Canfield},\ and\
  \citenamefont {Kaminski}}]{Wu2016}%
  \BibitemOpen
  \bibfield  {author} {\bibinfo {author} {\bibfnamefont {Y.}~\bibnamefont
  {Wu}}, \bibinfo {author} {\bibfnamefont {D.}~\bibnamefont {Mou}}, \bibinfo
  {author} {\bibfnamefont {N.~H.}\ \bibnamefont {Jo}}, \bibinfo {author}
  {\bibfnamefont {K.}~\bibnamefont {Sun}}, \bibinfo {author} {\bibfnamefont
  {L.}~\bibnamefont {Huang}}, \bibinfo {author} {\bibfnamefont {S.~L.}\
  \bibnamefont {Bud'ko}}, \bibinfo {author} {\bibfnamefont {P.~C.}\
  \bibnamefont {Canfield}}, \ and\ \bibinfo {author} {\bibfnamefont
  {A.}~\bibnamefont {Kaminski}},\ }\href {\doibase 10.1103/PhysRevB.94.121113}
  {\bibfield  {journal} {\bibinfo  {journal} {Phys. Rev. B}\ }\textbf {\bibinfo
  {volume} {94}},\ \bibinfo {pages} {121113 (R)} (\bibinfo {year}
  {2016})}\BibitemShut {NoStop}%
\bibitem [{\citenamefont {Wang}\ \emph
  {et~al.}(2016{\natexlab{b}})\citenamefont {Wang}, \citenamefont {Zhang},
  \citenamefont {Huang}, \citenamefont {Nie}, \citenamefont {Liu},
  \citenamefont {Liang}, \citenamefont {Zhang}, \citenamefont {Shen},
  \citenamefont {Liu}, \citenamefont {Hu}, \citenamefont {Ding}, \citenamefont
  {Liu}, \citenamefont {Hu}, \citenamefont {He}, \citenamefont {Zhao},
  \citenamefont {Yu}, \citenamefont {Hu}, \citenamefont {Wei}, \citenamefont
  {Mao}, \citenamefont {Shi}, \citenamefont {Jia}, \citenamefont {Zhang},
  \citenamefont {Zhang}, \citenamefont {Yang}, \citenamefont {Wang},
  \citenamefont {Peng}, \citenamefont {Weng}, \citenamefont {Dai},
  \citenamefont {Fang}, \citenamefont {Xu}, \citenamefont {Chen},\ and\
  \citenamefont {Zhou}}]{Wang2016b}%
  \BibitemOpen
  \bibfield  {author} {\bibinfo {author} {\bibfnamefont {C.}~\bibnamefont
  {Wang}}, \bibinfo {author} {\bibfnamefont {Y.}~\bibnamefont {Zhang}},
  \bibinfo {author} {\bibfnamefont {J.}~\bibnamefont {Huang}}, \bibinfo
  {author} {\bibfnamefont {S.}~\bibnamefont {Nie}}, \bibinfo {author}
  {\bibfnamefont {G.}~\bibnamefont {Liu}}, \bibinfo {author} {\bibfnamefont
  {A.}~\bibnamefont {Liang}}, \bibinfo {author} {\bibfnamefont
  {Y.}~\bibnamefont {Zhang}}, \bibinfo {author} {\bibfnamefont
  {B.}~\bibnamefont {Shen}}, \bibinfo {author} {\bibfnamefont {J.}~\bibnamefont
  {Liu}}, \bibinfo {author} {\bibfnamefont {C.}~\bibnamefont {Hu}}, \bibinfo
  {author} {\bibfnamefont {Y.}~\bibnamefont {Ding}}, \bibinfo {author}
  {\bibfnamefont {D.}~\bibnamefont {Liu}}, \bibinfo {author} {\bibfnamefont
  {Y.}~\bibnamefont {Hu}}, \bibinfo {author} {\bibfnamefont {S.}~\bibnamefont
  {He}}, \bibinfo {author} {\bibfnamefont {L.}~\bibnamefont {Zhao}}, \bibinfo
  {author} {\bibfnamefont {L.}~\bibnamefont {Yu}}, \bibinfo {author}
  {\bibfnamefont {J.}~\bibnamefont {Hu}}, \bibinfo {author} {\bibfnamefont
  {J.}~\bibnamefont {Wei}}, \bibinfo {author} {\bibfnamefont {Z.}~\bibnamefont
  {Mao}}, \bibinfo {author} {\bibfnamefont {Y.}~\bibnamefont {Shi}}, \bibinfo
  {author} {\bibfnamefont {X.}~\bibnamefont {Jia}}, \bibinfo {author}
  {\bibfnamefont {F.}~\bibnamefont {Zhang}}, \bibinfo {author} {\bibfnamefont
  {S.}~\bibnamefont {Zhang}}, \bibinfo {author} {\bibfnamefont
  {F.}~\bibnamefont {Yang}}, \bibinfo {author} {\bibfnamefont {Z.}~\bibnamefont
  {Wang}}, \bibinfo {author} {\bibfnamefont {Q.}~\bibnamefont {Peng}}, \bibinfo
  {author} {\bibfnamefont {H.}~\bibnamefont {Weng}}, \bibinfo {author}
  {\bibfnamefont {X.}~\bibnamefont {Dai}}, \bibinfo {author} {\bibfnamefont
  {Z.}~\bibnamefont {Fang}}, \bibinfo {author} {\bibfnamefont {Z.}~\bibnamefont
  {Xu}}, \bibinfo {author} {\bibfnamefont {C.}~\bibnamefont {Chen}}, \ and\
  \bibinfo {author} {\bibfnamefont {X.~J.}\ \bibnamefont {Zhou}},\ }\href
  {\doibase 10.1103/PhysRevB.94.241119} {\bibfield  {journal} {\bibinfo
  {journal} {Phys. Rev. B}\ }\textbf {\bibinfo {volume} {94}},\ \bibinfo
  {pages} {241119} (\bibinfo {year} {2016}{\natexlab{b}})}\BibitemShut
  {NoStop}%
\bibitem [{\citenamefont {Das}\ \emph {et~al.}(2016)\citenamefont {Das},
  \citenamefont {Sante}, \citenamefont {Vobornik}, \citenamefont {Fujii},
  \citenamefont {Okuda}, \citenamefont {Bruyer}, \citenamefont {Gyenis},
  \citenamefont {Feldman}, \citenamefont {Tao}, \citenamefont {Ciancio},
  \citenamefont {Rossi}, \citenamefont {Ali}, \citenamefont {Picozzi},
  \citenamefont {Yadzani}, \citenamefont {Panaccione},\ and\ \citenamefont
  {Cava}}]{Das2016}%
  \BibitemOpen
  \bibfield  {author} {\bibinfo {author} {\bibfnamefont {P.~K.}\ \bibnamefont
  {Das}}, \bibinfo {author} {\bibfnamefont {D.~D.}\ \bibnamefont {Sante}},
  \bibinfo {author} {\bibfnamefont {I.}~\bibnamefont {Vobornik}}, \bibinfo
  {author} {\bibfnamefont {J.}~\bibnamefont {Fujii}}, \bibinfo {author}
  {\bibfnamefont {T.}~\bibnamefont {Okuda}}, \bibinfo {author} {\bibfnamefont
  {E.}~\bibnamefont {Bruyer}}, \bibinfo {author} {\bibfnamefont
  {A.}~\bibnamefont {Gyenis}}, \bibinfo {author} {\bibfnamefont {B.~E.}\
  \bibnamefont {Feldman}}, \bibinfo {author} {\bibfnamefont {J.}~\bibnamefont
  {Tao}}, \bibinfo {author} {\bibfnamefont {R.}~\bibnamefont {Ciancio}},
  \bibinfo {author} {\bibfnamefont {G.}~\bibnamefont {Rossi}}, \bibinfo
  {author} {\bibfnamefont {M.~N.}\ \bibnamefont {Ali}}, \bibinfo {author}
  {\bibfnamefont {S.}~\bibnamefont {Picozzi}}, \bibinfo {author} {\bibfnamefont
  {A.}~\bibnamefont {Yadzani}}, \bibinfo {author} {\bibfnamefont
  {G.}~\bibnamefont {Panaccione}}, \ and\ \bibinfo {author} {\bibfnamefont
  {R.~J.}\ \bibnamefont {Cava}},\ }\href {\doibase 10.1038/ncomms10847}
  {\bibfield  {journal} {\bibinfo  {journal} {Nature Communications}\ }\textbf
  {\bibinfo {volume} {7}},\ \bibinfo {pages} {10847} (\bibinfo {year}
  {2016})}\BibitemShut {NoStop}%
\bibitem [{\citenamefont {Belopolski}\ \emph {et~al.}(2016)\citenamefont
  {Belopolski}, \citenamefont {Xu}, \citenamefont {Ishida}, \citenamefont
  {Pan}, \citenamefont {Yu}, \citenamefont {Sanchez}, \citenamefont {Zheng},
  \citenamefont {Neupane}, \citenamefont {Alidoust}, \citenamefont {Chang},
  \citenamefont {Chang}, \citenamefont {Wu}, \citenamefont {Bian},
  \citenamefont {Huang}, \citenamefont {Lee}, \citenamefont {Mou},
  \citenamefont {Huang}, \citenamefont {Song}, \citenamefont {Wang},
  \citenamefont {Wang}, \citenamefont {Yeh}, \citenamefont {Yao}, \citenamefont
  {Rault}, \citenamefont {Le~F\`evre}, \citenamefont {Bertran}, \citenamefont
  {Jeng}, \citenamefont {Kondo}, \citenamefont {Kaminski}, \citenamefont {Lin},
  \citenamefont {Liu}, \citenamefont {Song}, \citenamefont {Shin},\ and\
  \citenamefont {Hasan}}]{Belopolski2016}%
  \BibitemOpen
  \bibfield  {author} {\bibinfo {author} {\bibfnamefont {I.}~\bibnamefont
  {Belopolski}}, \bibinfo {author} {\bibfnamefont {S.-Y.}\ \bibnamefont {Xu}},
  \bibinfo {author} {\bibfnamefont {Y.}~\bibnamefont {Ishida}}, \bibinfo
  {author} {\bibfnamefont {X.}~\bibnamefont {Pan}}, \bibinfo {author}
  {\bibfnamefont {P.}~\bibnamefont {Yu}}, \bibinfo {author} {\bibfnamefont
  {D.~S.}\ \bibnamefont {Sanchez}}, \bibinfo {author} {\bibfnamefont
  {H.}~\bibnamefont {Zheng}}, \bibinfo {author} {\bibfnamefont
  {M.}~\bibnamefont {Neupane}}, \bibinfo {author} {\bibfnamefont
  {N.}~\bibnamefont {Alidoust}}, \bibinfo {author} {\bibfnamefont
  {G.}~\bibnamefont {Chang}}, \bibinfo {author} {\bibfnamefont {T.-R.}\
  \bibnamefont {Chang}}, \bibinfo {author} {\bibfnamefont {Y.}~\bibnamefont
  {Wu}}, \bibinfo {author} {\bibfnamefont {G.}~\bibnamefont {Bian}}, \bibinfo
  {author} {\bibfnamefont {S.-M.}\ \bibnamefont {Huang}}, \bibinfo {author}
  {\bibfnamefont {C.-C.}\ \bibnamefont {Lee}}, \bibinfo {author} {\bibfnamefont
  {D.}~\bibnamefont {Mou}}, \bibinfo {author} {\bibfnamefont {L.}~\bibnamefont
  {Huang}}, \bibinfo {author} {\bibfnamefont {Y.}~\bibnamefont {Song}},
  \bibinfo {author} {\bibfnamefont {B.}~\bibnamefont {Wang}}, \bibinfo {author}
  {\bibfnamefont {G.}~\bibnamefont {Wang}}, \bibinfo {author} {\bibfnamefont
  {Y.-W.}\ \bibnamefont {Yeh}}, \bibinfo {author} {\bibfnamefont
  {N.}~\bibnamefont {Yao}}, \bibinfo {author} {\bibfnamefont {J.~E.}\
  \bibnamefont {Rault}}, \bibinfo {author} {\bibfnamefont {P.}~\bibnamefont
  {Le~F\`evre}}, \bibinfo {author} {\bibfnamefont {F.~m.~c.}\ \bibnamefont
  {Bertran}}, \bibinfo {author} {\bibfnamefont {H.-T.}\ \bibnamefont {Jeng}},
  \bibinfo {author} {\bibfnamefont {T.}~\bibnamefont {Kondo}}, \bibinfo
  {author} {\bibfnamefont {A.}~\bibnamefont {Kaminski}}, \bibinfo {author}
  {\bibfnamefont {H.}~\bibnamefont {Lin}}, \bibinfo {author} {\bibfnamefont
  {Z.}~\bibnamefont {Liu}}, \bibinfo {author} {\bibfnamefont {F.}~\bibnamefont
  {Song}}, \bibinfo {author} {\bibfnamefont {S.}~\bibnamefont {Shin}}, \ and\
  \bibinfo {author} {\bibfnamefont {M.~Z.}\ \bibnamefont {Hasan}},\ }\href
  {\doibase 10.1103/PhysRevB.94.085127} {\bibfield  {journal} {\bibinfo
  {journal} {Phys. Rev. B}\ }\textbf {\bibinfo {volume} {94}},\ \bibinfo
  {pages} {085127} (\bibinfo {year} {2016})}\BibitemShut {NoStop}%
\bibitem [{\citenamefont {Feng}\ \emph {et~al.}(2016)\citenamefont {Feng},
  \citenamefont {Chan}, \citenamefont {Feng}, \citenamefont {Liu},
  \citenamefont {Chou}, \citenamefont {Kuroda}, \citenamefont {Yaji},
  \citenamefont {Harasawa}, \citenamefont {Moras}, \citenamefont {Barinov},
  \citenamefont {Malaeb}, \citenamefont {Bareille}, \citenamefont {Kondo},
  \citenamefont {Shin}, \citenamefont {Komori}, \citenamefont {Chiang},
  \citenamefont {Shi},\ and\ \citenamefont {Matsuda}}]{Feng2016}%
  \BibitemOpen
  \bibfield  {author} {\bibinfo {author} {\bibfnamefont {B.}~\bibnamefont
  {Feng}}, \bibinfo {author} {\bibfnamefont {Y.-H.}\ \bibnamefont {Chan}},
  \bibinfo {author} {\bibfnamefont {Y.}~\bibnamefont {Feng}}, \bibinfo {author}
  {\bibfnamefont {R.-Y.}\ \bibnamefont {Liu}}, \bibinfo {author} {\bibfnamefont
  {M.-Y.}\ \bibnamefont {Chou}}, \bibinfo {author} {\bibfnamefont
  {K.}~\bibnamefont {Kuroda}}, \bibinfo {author} {\bibfnamefont
  {K.}~\bibnamefont {Yaji}}, \bibinfo {author} {\bibfnamefont {A.}~\bibnamefont
  {Harasawa}}, \bibinfo {author} {\bibfnamefont {P.}~\bibnamefont {Moras}},
  \bibinfo {author} {\bibfnamefont {A.}~\bibnamefont {Barinov}}, \bibinfo
  {author} {\bibfnamefont {W.}~\bibnamefont {Malaeb}}, \bibinfo {author}
  {\bibfnamefont {C.}~\bibnamefont {Bareille}}, \bibinfo {author}
  {\bibfnamefont {T.}~\bibnamefont {Kondo}}, \bibinfo {author} {\bibfnamefont
  {S.}~\bibnamefont {Shin}}, \bibinfo {author} {\bibfnamefont {F.}~\bibnamefont
  {Komori}}, \bibinfo {author} {\bibfnamefont {T.-C.}\ \bibnamefont {Chiang}},
  \bibinfo {author} {\bibfnamefont {Y.}~\bibnamefont {Shi}}, \ and\ \bibinfo
  {author} {\bibfnamefont {I.}~\bibnamefont {Matsuda}},\ }\href {\doibase
  10.1103/PhysRevB.94.195134} {\bibfield  {journal} {\bibinfo  {journal} {Phys.
  Rev. B}\ }\textbf {\bibinfo {volume} {94}},\ \bibinfo {pages} {195134}
  (\bibinfo {year} {2016})}\BibitemShut {NoStop}%
\bibitem [{\citenamefont {S\'anchez-Barriga}\ \emph {et~al.}(2016)\citenamefont
  {S\'anchez-Barriga}, \citenamefont {Vergniory}, \citenamefont {Evtushinsky},
  \citenamefont {Aguilera}, \citenamefont {Varykhalov}, \citenamefont
  {Bl\"ugel},\ and\ \citenamefont {Rader}}]{Sanchez-Barriga2016a}%
  \BibitemOpen
  \bibfield  {author} {\bibinfo {author} {\bibfnamefont {J.}~\bibnamefont
  {S\'anchez-Barriga}}, \bibinfo {author} {\bibfnamefont {M.~G.}\ \bibnamefont
  {Vergniory}}, \bibinfo {author} {\bibfnamefont {D.}~\bibnamefont
  {Evtushinsky}}, \bibinfo {author} {\bibfnamefont {I.}~\bibnamefont
  {Aguilera}}, \bibinfo {author} {\bibfnamefont {A.}~\bibnamefont
  {Varykhalov}}, \bibinfo {author} {\bibfnamefont {S.}~\bibnamefont
  {Bl\"ugel}}, \ and\ \bibinfo {author} {\bibfnamefont {O.}~\bibnamefont
  {Rader}},\ }\href {\doibase 10.1103/PhysRevB.94.161401} {\bibfield  {journal}
  {\bibinfo  {journal} {Phys. Rev. B}\ }\textbf {\bibinfo {volume} {94}},\
  \bibinfo {pages} {161401 (R)} (\bibinfo {year} {2016})}\BibitemShut {NoStop}%
\bibitem [{\citenamefont {Brown}(1966)}]{Brown1966}%
  \BibitemOpen
  \bibfield  {author} {\bibinfo {author} {\bibfnamefont {B.~E.}\ \bibnamefont
  {Brown}},\ }\href {\doibase 10.1107/s0365110x66000513} {\bibfield  {journal}
  {\bibinfo  {journal} {Acta Crystallographica}\ }\textbf {\bibinfo {volume}
  {20}},\ \bibinfo {pages} {268} (\bibinfo {year} {1966})}\BibitemShut
  {NoStop}%
\bibitem [{\citenamefont {Perdew}\ \emph {et~al.}(1996)\citenamefont {Perdew},
  \citenamefont {Burke},\ and\ \citenamefont {Ernzerhof}}]{Perdew1996}%
  \BibitemOpen
  \bibfield  {author} {\bibinfo {author} {\bibfnamefont {J.~P.}\ \bibnamefont
  {Perdew}}, \bibinfo {author} {\bibfnamefont {K.}~\bibnamefont {Burke}}, \
  and\ \bibinfo {author} {\bibfnamefont {M.}~\bibnamefont {Ernzerhof}},\ }\href
  {\doibase 10.1103/PhysRevLett.77.3865} {\bibfield  {journal} {\bibinfo
  {journal} {Phys. Rev. Lett.}\ }\textbf {\bibinfo {volume} {77}},\ \bibinfo
  {pages} {3865} (\bibinfo {year} {1996})}\BibitemShut {NoStop}%
\bibitem [{\citenamefont {Giannozzi}\ \emph {et~al.}(2009)\citenamefont
  {Giannozzi}, \citenamefont {Baroni}, \citenamefont {Bonini}, \citenamefont
  {Calandra}, \citenamefont {Car}, \citenamefont {Cavazzoni}, \citenamefont
  {Ceresoli}, \citenamefont {Chiarotti}, \citenamefont {Cococcioni},
  \citenamefont {Dabo}, \citenamefont {{Dal Corso}}, \citenamefont
  {de~Gironcoli}, \citenamefont {Fabris}, \citenamefont {Fratesi},
  \citenamefont {Gebauer}, \citenamefont {Gerstmann}, \citenamefont
  {Gougoussis}, \citenamefont {Kokalj}, \citenamefont {Lazzeri}, \citenamefont
  {Martin-Samos}, \citenamefont {Marzari}, \citenamefont {Mauri}, \citenamefont
  {Mazzarello}, \citenamefont {Paolini}, \citenamefont {Pasquarello},
  \citenamefont {Paulatto}, \citenamefont {Sbraccia}, \citenamefont {Scandolo},
  \citenamefont {Sclauzero}, \citenamefont {Seitsonen}, \citenamefont
  {Smogunov}, \citenamefont {Umari},\ and\ \citenamefont
  {Wentzcovitch}}]{QE-2009}%
  \BibitemOpen
  \bibfield  {author} {\bibinfo {author} {\bibfnamefont {P.}~\bibnamefont
  {Giannozzi}}, \bibinfo {author} {\bibfnamefont {S.}~\bibnamefont {Baroni}},
  \bibinfo {author} {\bibfnamefont {N.}~\bibnamefont {Bonini}}, \bibinfo
  {author} {\bibfnamefont {M.}~\bibnamefont {Calandra}}, \bibinfo {author}
  {\bibfnamefont {R.}~\bibnamefont {Car}}, \bibinfo {author} {\bibfnamefont
  {C.}~\bibnamefont {Cavazzoni}}, \bibinfo {author} {\bibfnamefont
  {D.}~\bibnamefont {Ceresoli}}, \bibinfo {author} {\bibfnamefont {G.~L.}\
  \bibnamefont {Chiarotti}}, \bibinfo {author} {\bibfnamefont {M.}~\bibnamefont
  {Cococcioni}}, \bibinfo {author} {\bibfnamefont {I.}~\bibnamefont {Dabo}},
  \bibinfo {author} {\bibfnamefont {A.}~\bibnamefont {{Dal Corso}}}, \bibinfo
  {author} {\bibfnamefont {S.}~\bibnamefont {de~Gironcoli}}, \bibinfo {author}
  {\bibfnamefont {S.}~\bibnamefont {Fabris}}, \bibinfo {author} {\bibfnamefont
  {G.}~\bibnamefont {Fratesi}}, \bibinfo {author} {\bibfnamefont
  {R.}~\bibnamefont {Gebauer}}, \bibinfo {author} {\bibfnamefont
  {U.}~\bibnamefont {Gerstmann}}, \bibinfo {author} {\bibfnamefont
  {C.}~\bibnamefont {Gougoussis}}, \bibinfo {author} {\bibfnamefont
  {A.}~\bibnamefont {Kokalj}}, \bibinfo {author} {\bibfnamefont
  {M.}~\bibnamefont {Lazzeri}}, \bibinfo {author} {\bibfnamefont
  {L.}~\bibnamefont {Martin-Samos}}, \bibinfo {author} {\bibfnamefont
  {N.}~\bibnamefont {Marzari}}, \bibinfo {author} {\bibfnamefont
  {F.}~\bibnamefont {Mauri}}, \bibinfo {author} {\bibfnamefont
  {R.}~\bibnamefont {Mazzarello}}, \bibinfo {author} {\bibfnamefont
  {S.}~\bibnamefont {Paolini}}, \bibinfo {author} {\bibfnamefont
  {A.}~\bibnamefont {Pasquarello}}, \bibinfo {author} {\bibfnamefont
  {L.}~\bibnamefont {Paulatto}}, \bibinfo {author} {\bibfnamefont
  {C.}~\bibnamefont {Sbraccia}}, \bibinfo {author} {\bibfnamefont
  {S.}~\bibnamefont {Scandolo}}, \bibinfo {author} {\bibfnamefont
  {G.}~\bibnamefont {Sclauzero}}, \bibinfo {author} {\bibfnamefont {A.~P.}\
  \bibnamefont {Seitsonen}}, \bibinfo {author} {\bibfnamefont {A.}~\bibnamefont
  {Smogunov}}, \bibinfo {author} {\bibfnamefont {P.}~\bibnamefont {Umari}}, \
  and\ \bibinfo {author} {\bibfnamefont {R.~M.}\ \bibnamefont {Wentzcovitch}},\
  }\href {http://www.quantum-espresso.org} {\bibfield  {journal} {\bibinfo
  {journal} {J. Phys.: Condens. Matter}\ }\textbf {\bibinfo {volume} {21}},\
  \bibinfo {pages} {395502 (19pp)} (\bibinfo {year} {2009})}\BibitemShut
  {NoStop}%
\bibitem [{\citenamefont {{Di Sante}}\ \emph {et~al.}(2017)\citenamefont {{Di
  Sante}}, \citenamefont {{Das}}, \citenamefont {{Bigi}}, \citenamefont
  {{Erg{\"o}nenc}}, \citenamefont {{G{\"u}rtler}}, \citenamefont {{Krieger}},
  \citenamefont {{Schmitt}}, \citenamefont {{Ali}}, \citenamefont {{Rossi}},
  \citenamefont {{Thomale}}, \citenamefont {{Franchini}}, \citenamefont
  {{Picozzi}}, \citenamefont {{Fujii}}, \citenamefont {{Strocov}},
  \citenamefont {{Sangiovanni}}, \citenamefont {{Vobornik}}, \citenamefont
  {{Cava}},\ and\ \citenamefont {{Panaccione}}}]{DiSante2017}%
  \BibitemOpen
  \bibfield  {author} {\bibinfo {author} {\bibfnamefont {D.}~\bibnamefont {{Di
  Sante}}}, \bibinfo {author} {\bibfnamefont {P.~K.}\ \bibnamefont {{Das}}},
  \bibinfo {author} {\bibfnamefont {C.}~\bibnamefont {{Bigi}}}, \bibinfo
  {author} {\bibfnamefont {Z.}~\bibnamefont {{Erg{\"o}nenc}}}, \bibinfo
  {author} {\bibfnamefont {N.}~\bibnamefont {{G{\"u}rtler}}}, \bibinfo {author}
  {\bibfnamefont {J.~A.}\ \bibnamefont {{Krieger}}}, \bibinfo {author}
  {\bibfnamefont {T.}~\bibnamefont {{Schmitt}}}, \bibinfo {author}
  {\bibfnamefont {M.~N.}\ \bibnamefont {{Ali}}}, \bibinfo {author}
  {\bibfnamefont {G.}~\bibnamefont {{Rossi}}}, \bibinfo {author} {\bibfnamefont
  {R.}~\bibnamefont {{Thomale}}}, \bibinfo {author} {\bibfnamefont
  {C.}~\bibnamefont {{Franchini}}}, \bibinfo {author} {\bibfnamefont
  {S.}~\bibnamefont {{Picozzi}}}, \bibinfo {author} {\bibfnamefont
  {J.}~\bibnamefont {{Fujii}}}, \bibinfo {author} {\bibfnamefont {V.~N.}\
  \bibnamefont {{Strocov}}}, \bibinfo {author} {\bibfnamefont {G.}~\bibnamefont
  {{Sangiovanni}}}, \bibinfo {author} {\bibfnamefont {I.}~\bibnamefont
  {{Vobornik}}}, \bibinfo {author} {\bibfnamefont {R.~J.}\ \bibnamefont
  {{Cava}}}, \ and\ \bibinfo {author} {\bibfnamefont {G.}~\bibnamefont
  {{Panaccione}}},\ }\href@noop {} {\bibfield  {journal} {\bibinfo  {journal}
  {ArXiv:1702.05322}\ } (\bibinfo {year} {2017})},\ \Eprint
  {http://arxiv.org/abs/1702.05322} {arXiv:1702.05322 [cond-mat.mtrl-sci]}
  \BibitemShut {NoStop}%
\bibitem [{\citenamefont {Dawson}\ and\ \citenamefont
  {Bullett}(1987)}]{Dawson1987}%
  \BibitemOpen
  \bibfield  {author} {\bibinfo {author} {\bibfnamefont {W.~G.}\ \bibnamefont
  {Dawson}}\ and\ \bibinfo {author} {\bibfnamefont {D.~W.}\ \bibnamefont
  {Bullett}},\ }\href {\doibase 10.1088/0022-3719/20/36/017} {\bibfield
  {journal} {\bibinfo  {journal} {Journal of Physics C: Solid State Physics}\
  }\textbf {\bibinfo {volume} {20}},\ \bibinfo {pages} {6159} (\bibinfo {year}
  {1987})}\BibitemShut {NoStop}%
\bibitem [{\citenamefont {Augustin}\ \emph {et~al.}(2000)\citenamefont
  {Augustin}, \citenamefont {Eyert}, \citenamefont {Böker}, \citenamefont
  {Frentrup}, \citenamefont {Dwelk}, \citenamefont {Janowitz},\ and\
  \citenamefont {Manzke}}]{Augustin2000}%
  \BibitemOpen
  \bibfield  {author} {\bibinfo {author} {\bibfnamefont {J.}~\bibnamefont
  {Augustin}}, \bibinfo {author} {\bibfnamefont {V.}~\bibnamefont {Eyert}},
  \bibinfo {author} {\bibfnamefont {T.}~\bibnamefont {Böker}}, \bibinfo
  {author} {\bibfnamefont {W.}~\bibnamefont {Frentrup}}, \bibinfo {author}
  {\bibfnamefont {H.}~\bibnamefont {Dwelk}}, \bibinfo {author} {\bibfnamefont
  {C.}~\bibnamefont {Janowitz}}, \ and\ \bibinfo {author} {\bibfnamefont
  {R.}~\bibnamefont {Manzke}},\ }\href {\doibase 10.1103/physrevb.62.10812}
  {\bibfield  {journal} {\bibinfo  {journal} {Physical Review B}\ }\textbf
  {\bibinfo {volume} {62}},\ \bibinfo {pages} {10812} (\bibinfo {year}
  {2000})}\BibitemShut {NoStop}%
\bibitem [{\citenamefont {Crepaldi}\ \emph {et~al.}(2017)\citenamefont
  {Crepaldi}, \citenamefont {Aut\`es}, \citenamefont {Sterzi}, \citenamefont
  {Manzoni}, \citenamefont {Zacchigna}, \citenamefont {Cilento}, \citenamefont
  {Vobornik}, \citenamefont {Fujii}, \citenamefont {Bugnon}, \citenamefont
  {Magrez}, \citenamefont {Berger}, \citenamefont {Parmigiani}, \citenamefont
  {Yazyev},\ and\ \citenamefont {Grioni}}]{Crepaldi2017}%
  \BibitemOpen
  \bibfield  {author} {\bibinfo {author} {\bibfnamefont {A.}~\bibnamefont
  {Crepaldi}}, \bibinfo {author} {\bibfnamefont {G.}~\bibnamefont {Aut\`es}},
  \bibinfo {author} {\bibfnamefont {A.}~\bibnamefont {Sterzi}}, \bibinfo
  {author} {\bibfnamefont {G.}~\bibnamefont {Manzoni}}, \bibinfo {author}
  {\bibfnamefont {M.}~\bibnamefont {Zacchigna}}, \bibinfo {author}
  {\bibfnamefont {F.}~\bibnamefont {Cilento}}, \bibinfo {author} {\bibfnamefont
  {I.}~\bibnamefont {Vobornik}}, \bibinfo {author} {\bibfnamefont
  {J.}~\bibnamefont {Fujii}}, \bibinfo {author} {\bibfnamefont
  {P.}~\bibnamefont {Bugnon}}, \bibinfo {author} {\bibfnamefont
  {A.}~\bibnamefont {Magrez}}, \bibinfo {author} {\bibfnamefont
  {H.}~\bibnamefont {Berger}}, \bibinfo {author} {\bibfnamefont
  {F.}~\bibnamefont {Parmigiani}}, \bibinfo {author} {\bibfnamefont {O.~V.}\
  \bibnamefont {Yazyev}}, \ and\ \bibinfo {author} {\bibfnamefont
  {M.}~\bibnamefont {Grioni}},\ }\href {\doibase 10.1103/PhysRevB.95.041408}
  {\bibfield  {journal} {\bibinfo  {journal} {Phys. Rev. B}\ }\textbf {\bibinfo
  {volume} {95}},\ \bibinfo {pages} {041408} (\bibinfo {year}
  {2017})}\BibitemShut {NoStop}%
\bibitem [{\citenamefont {Kawahara}\ \emph {et~al.}(2017)\citenamefont
  {Kawahara}, \citenamefont {Ni}, \citenamefont {Arafune}, \citenamefont
  {Shirasawa}, \citenamefont {Lin}, \citenamefont {Minamitani}, \citenamefont
  {Watanabe}, \citenamefont {Kawai},\ and\ \citenamefont
  {Takagi}}]{Kawahara2017}%
  \BibitemOpen
  \bibfield  {author} {\bibinfo {author} {\bibfnamefont {K.}~\bibnamefont
  {Kawahara}}, \bibinfo {author} {\bibfnamefont {Z.}~\bibnamefont {Ni}},
  \bibinfo {author} {\bibfnamefont {R.}~\bibnamefont {Arafune}}, \bibinfo
  {author} {\bibfnamefont {T.}~\bibnamefont {Shirasawa}}, \bibinfo {author}
  {\bibfnamefont {C.-L.}\ \bibnamefont {Lin}}, \bibinfo {author} {\bibfnamefont
  {E.}~\bibnamefont {Minamitani}}, \bibinfo {author} {\bibfnamefont
  {S.}~\bibnamefont {Watanabe}}, \bibinfo {author} {\bibfnamefont
  {M.}~\bibnamefont {Kawai}}, \ and\ \bibinfo {author} {\bibfnamefont
  {N.}~\bibnamefont {Takagi}},\ }\href {\doibase 10.7567/apex.10.045702}
  {\bibfield  {journal} {\bibinfo  {journal} {Applied Physics Express}\
  }\textbf {\bibinfo {volume} {10}},\ \bibinfo {pages} {045702} (\bibinfo
  {year} {2017})}\BibitemShut {NoStop}%
\bibitem [{\citenamefont {Luo}\ \emph {et~al.}(2015)\citenamefont {Luo},
  \citenamefont {Li}, \citenamefont {Dai}, \citenamefont {Miao}, \citenamefont
  {Shi}, \citenamefont {Ding}, \citenamefont {Taylor}, \citenamefont
  {Yarotski}, \citenamefont {Prasankumar},\ and\ \citenamefont
  {Thompson}}]{Luo2015}%
  \BibitemOpen
  \bibfield  {author} {\bibinfo {author} {\bibfnamefont {Y.}~\bibnamefont
  {Luo}}, \bibinfo {author} {\bibfnamefont {H.}~\bibnamefont {Li}}, \bibinfo
  {author} {\bibfnamefont {Y.~M.}\ \bibnamefont {Dai}}, \bibinfo {author}
  {\bibfnamefont {H.}~\bibnamefont {Miao}}, \bibinfo {author} {\bibfnamefont
  {Y.~G.}\ \bibnamefont {Shi}}, \bibinfo {author} {\bibfnamefont
  {H.}~\bibnamefont {Ding}}, \bibinfo {author} {\bibfnamefont {A.~J.}\
  \bibnamefont {Taylor}}, \bibinfo {author} {\bibfnamefont {D.~A.}\
  \bibnamefont {Yarotski}}, \bibinfo {author} {\bibfnamefont {R.~P.}\
  \bibnamefont {Prasankumar}}, \ and\ \bibinfo {author} {\bibfnamefont {J.~D.}\
  \bibnamefont {Thompson}},\ }\href {\doibase 10.1063/1.4935240} {\bibfield
  {journal} {\bibinfo  {journal} {Applied Physics Letters}\ }\textbf {\bibinfo
  {volume} {107}},\ \bibinfo {pages} {182411} (\bibinfo {year}
  {2015})}\BibitemShut {NoStop}%
\bibitem [{\citenamefont {Wang}\ \emph {et~al.}(2015)\citenamefont {Wang},
  \citenamefont {Thoutam}, \citenamefont {Xiao}, \citenamefont {Hu},
  \citenamefont {Das}, \citenamefont {Mao}, \citenamefont {Wei}, \citenamefont
  {Divan}, \citenamefont {Luican-Mayer}, \citenamefont {Crabtree},\ and\
  \citenamefont {Kwok}}]{Wang2015b}%
  \BibitemOpen
  \bibfield  {author} {\bibinfo {author} {\bibfnamefont {Y.~L.}\ \bibnamefont
  {Wang}}, \bibinfo {author} {\bibfnamefont {L.~R.}\ \bibnamefont {Thoutam}},
  \bibinfo {author} {\bibfnamefont {Z.~L.}\ \bibnamefont {Xiao}}, \bibinfo
  {author} {\bibfnamefont {J.}~\bibnamefont {Hu}}, \bibinfo {author}
  {\bibfnamefont {S.}~\bibnamefont {Das}}, \bibinfo {author} {\bibfnamefont
  {Z.~Q.}\ \bibnamefont {Mao}}, \bibinfo {author} {\bibfnamefont
  {J.}~\bibnamefont {Wei}}, \bibinfo {author} {\bibfnamefont {R.}~\bibnamefont
  {Divan}}, \bibinfo {author} {\bibfnamefont {A.}~\bibnamefont {Luican-Mayer}},
  \bibinfo {author} {\bibfnamefont {G.~W.}\ \bibnamefont {Crabtree}}, \ and\
  \bibinfo {author} {\bibfnamefont {W.~K.}\ \bibnamefont {Kwok}},\ }\href
  {\doibase 10.1103/PhysRevB.92.180402} {\bibfield  {journal} {\bibinfo
  {journal} {Phys. Rev. B}\ }\textbf {\bibinfo {volume} {92}},\ \bibinfo
  {pages} {180402} (\bibinfo {year} {2015})}\BibitemShut {NoStop}%
\end{thebibliography}%

\end{document}